\documentclass[%
reprint,
amsmath,amssymb,
aps,
]{revtex4-2}
\usepackage{graphicx}
\usepackage{dcolumn}
\usepackage{bm}
\usepackage{physics}
\usepackage{amsmath}
\usepackage{natbib}
\usepackage{booktabs}
\usepackage{array}
\usepackage{float}
\usepackage{multirow}
\usepackage[colorlinks,linkcolor=blue,urlcolor=blue,anchorcolor=blue,citecolor=blue]{hyperref}
\usepackage{placeins}
\renewcommand{\figurename}{Figure}

\begin{document}
	
	\preprint{}
	
	\title{Probing the Anomalous Hall Transport and Magnetic Reversal of Chiral-Lattice Antiferromagnet Co$_{1/3}$NbS$_2$}
	
	
	\author{Pingfan Gu$^{1}$}
	\altaffiliation{These authors contributed equally to this work.}
	\author{Yuxuan Peng$^{1}$}
	\altaffiliation{These authors contributed equally to this work.}
	\author{Shiqi Yang$^{1,2}$}
	\author{Huan Wang$^{3}$} 
	\author{Shenyong Ye$^{1}$}
	\author{Hanwen Wang$^{4}$}
        \author{Yanping Li$^{1}$}
	\author{Tianlong Xia$^{3,5,6}$}
	\email{tlxia@ruc.edu.cn}
	\author{Jinbo Yang$^{1}$}
	\email{jbyang@pku.edu.cn}
	\author{Yu Ye$^{1,7,8}$}
	\email{ye$_$yu@pku.edu.cn}

	\affiliation{%
        $^{1}$State Key Laboratory for Artificial Microstructure $\rm{\&}$ Mesoscopic Physics and Frontiers Science Center for Nano-Optoelectronics, School of Physics, Peking University, Beijing 100871, China\\
        $^{2}$ Academy for Advanced Interdisciplinary Studies, Peking University, Beijing 100871, China\\
        $^{3}$Department of Physics and Beijing Key Laboratory of Optoelectronic Functional Materials and Micro-Nano Devices, Renmin University of China, Beijing 100872, China\\
        $^{4}$Shenyang National Laboratory for Materials Science, Institute of Metal Research, Chinese Academy of Sciences, Shenyang 110016, China\\
        $^{5}$Key Laboratory of Quantum State Construction and Manipulation (Ministry of Education), Renmin University of China, Beijing, 100872, China\\
        $^{6}$Laboratory for Neutron Scattering, Renmin University of China, Beijing 100872, China\\
        $^{7}$Collaboration International Center of Quantum Matter, Beijing 100871, China\\
        $^{8}$Liaoning Academy of Materials, Shenyang, 110167,  China\\
        }
	%
	
	
	
	\begin{abstract}
		Antiferromagnets exhibiting giant anomalous Hall effect (AHE) and anomalous Nernst effect (ANE) have recently aroused broad interest, not only for their potential applications in future electronic devices, but also because of the rich physics arising from the Berry curvature near the Fermi level. $\rm{Co_{1/3}NbS_2}$, by intercalating $\rm{Co^{2+}}$ ions between $\rm{NbS_2}$ layers, is a quasi-two-dimensional layered antiferromagnet with a chiral lattice. A large AHE has been observed in $\rm{Co_{1/3}NbS_2}$, but its origin is under debate. In this letter, we report the large AHE and ANE in exfoliated $\rm{Co_{1/3}NbS_2}$ flakes. By analyzing the thermoelectric data \textit{via} the Mott relation, we determined that the observed large AHE and ANE primarily result from the intrinsic Berry curvature. We also observed the magnetic domains in $\rm{Co_{1/3}NbS_2}$ by reflective magnetic circular dichroism measurements. Combined with electrical transport measurements, we confirmed that the magnetic reversal in $\rm{Co_{1/3}NbS_2}$ is determined by domain wall motion, and the critical field ($H_c$) exhibits a memory effect of consecutive magnetic sweeps. Our work provides insight into the topological properties of $\rm{Co_{1/3}NbS_2}$ and paves the way to studying the spin configuration and magnetic domain dynamics in this fascinating antiferromagnet.
	\end{abstract}
	
	
	\maketitle
	\section{Introduction}
	The spontaneous Hall effect, one of the most preferred methods for reading out spin polarization in metallic ferromagnets, has been discovered for over a century and is considered a hallmark of long-range ferromagnetism. However, it was only recently that scientists recognized the physical origin of the anomalous Hall value, specifically the Berry curvature and broken time-reversal symmetry\cite{nagaosa2010anomalous}. The profound understanding naturally predicts anomalous Hall effect (AHE) in antiferromagnets with nontrivial spin textures\cite{chen2014anomalous,martin2008itinerant,shindou2001orbital,taguchi2001spin,ohgushi2000spin}, where zero net magnetization is a tight demand for realizing next-generation ultra-compact spintronic devices. Soon after, AHE was observed experimentally in antiferromagnetic manganese compounds such as $\rm{Mn_3Sn}$\cite{nakatsuji2015large}, $\rm{Mn_3Ge}$\cite{nayak2016large}, $\rm{Mn_3Ga}$\cite{liu2017transition} and $\rm{Mn_5Si_3}$\cite{surgers2014large}. In these hexagonal materials, the cluster magnetic octupole serves as the order parameter to break the time-reversal symmetry and stabilize the Weyl fermions near the Fermi level\cite{chen2021anomalous}, which is also reversible by an external field resembling the magnetic moment.\par
	
	Even more fascinating, is the extension of AHE studies to two-dimensional (2D) antiferromagnets to explore topologically nontrivial energy bands\cite{deng2020quantum} and spin textures\cite{yang2021odd}. Recently, the intercalating 3$d$ magnetic atoms between layers of van der Waals transition metal dichalcogenide (TMDC) has opened up a new route to construct quasi-2D metals with divergent spin configurations and modified band structures. For example, a one-dimensional chiral magnetic soliton lattice is reported in $\rm{Cr_{1/3}NbS_2}$\cite{togawa2012chiral}. Electrical switching\cite{nair2020electrical} and exchange bias\cite{maniv2021exchange} are reported in $\rm{Fe_{1/3}NbS_2}$, where antiferromagnetic and frustrated spin-glass orders coexist and are evidenced to be coupled\cite{maniv2021exchange}. Among these intercalated compounds, $\rm{Co_{x}NbS_2}$ is well-known for its unexpectedly large AHE value, approaching the quantized conductance value of $e^2/h$ per layer\cite{ghimire2018large,tenasini2020giant}. The intercalated $\rm{Co^{2+}}$ cations serve to break the inversion symmetry and shift the Fermi level of $\rm{NbS_2}$, resulting in extra electronic bands that possibly contribute to AHE\cite{tanaka2022large,yang2022visualizing,popvcevic2022role}. The anomalous transport behavior is, therefore, sensitively determined by the stoichiometric composition $x$ with an idealized value of 1/3\cite{mangelsen2021interplay}. \par
	
	Despite extensive studies on this promising material, the spin configuration of $\rm{Co_{1/3}NbS_2}$ and the origin of AHE remain controversial. The earliest neutron diffraction results were fitted with multi-domain structures of collinear single $q$, with six symmetry-related in-plane $q$ sharing equal weights\cite{parkin1983magnetic}. However, this collinear structure is generally incompatible with the large AHE observed afterward. To account for the AHE value, Zhang et al. attributed it to the large hidden Berry curvature due to the chiral Dirac-like fermions \cite{zhang2023chiral}. On the other hand, {\v{S}}mejkal, et al. proposed a picture of crystal Hall effect \cite{vsmejkal2020crystal}, where magnetic orbitals rather than ordered spins break time-reversal symmetry, giving rise to spontaneous Hall signals. Lu, et al. further pointed out that the spins of $\rm{Co_{1/3}NbS_2}$ in the $bc$-plane are nearly anti-parallel, but the spins in adjacent $ab$-planes are alternately titled\cite{lu2022understanding}, which can quantitatively reproduce the AHE value. Meanwhile, Tenasini, et al.\cite{tenasini2020giant} validated the previous neutron scattering results but proposed a noncoplanar single-domain multi-$q$  magnetic structure to fit the data. Uncompensated Berry curvature in noncoplanar structure can explain the large AHE and is supported by first-principles calculations\cite{park2022first,heinonen2022magnetic}. Furthermore, Takagi et al. performed polarized neutron scattering experiments and determined an all-in-all-out type non-coplanar magnetic order, and the AHE can be explained in terms of the topological Hall effect originating from a fictitious magnetic field associated with the scalar spin chirality\cite{takagi2023spontaneous}.\par

    Taken together, the key factor hindering the determination of the actual magnetic order is the experimentally observed weak spontaneous net magnetization $\Delta M$, which is permissible in both configurations. The presence of $\Delta M$ is required for AHE to appear in a single-$q$ multi-domain configuration, but $\Delta M$ can theoretically be absent in a multi-$q$ single-domain configuration. Consequently, it is difficult to explicitly understand the AHE in $\rm{Co_{1/3}NbS_2}$ only by unraveling the complicated spin texture. In this work, we step over the specific spin configuration, but demonstrate that uniquely sensitive transport and optical measurements may provide even more crucial information. We performed magnetic, transport, thermoelectric, and reflective magnetic circular dichroism (RMCD) measurements on $\rm{Co_{1/3}NbS_2}$. By analyzing thermoelectric and transport data \textit{via} the Mott relation, we evidenced that the large AHE in $\rm{Co_{1/3}NbS_2}$ results from the intrinsic Berry curvature. In the exfoliated $\rm{Co_{1/3}NbS_2}$ below the Néel temperature a large RMCD signal comparable to that of ferromagnetic materials appears, verifying the large Berry curvature from the optical standpoint. Through RMCD mapping, we directly observed magnetic domains in $\rm{Co_{1/3}NbS_2}$, which will provide crucial information for the determination of the exact magnetic order. In combination with transport measurements, we confirmed that the magnetic reversal in $\rm{Co_{1/3}NbS_2}$ is dominated by domain wall motion, and the critical field ($H_c$) exhibits a memory effect of consecutive magnetic field sweeps. The underlying mechanism of this memory effect remains unclear but is certainly related to domain wall motion energy. Our work provides a more phenomenological and clearer understanding of the anomalous Hall and the magnetic reversal behavior in $\rm{Co_{1/3}NbS_2}$, which is essential for future applications and detailed studies of this enigmatic material.\par

	\section{Results and Discussion}
	
	\begin{figure}
		\centering
		\includegraphics[width=1.0\linewidth]{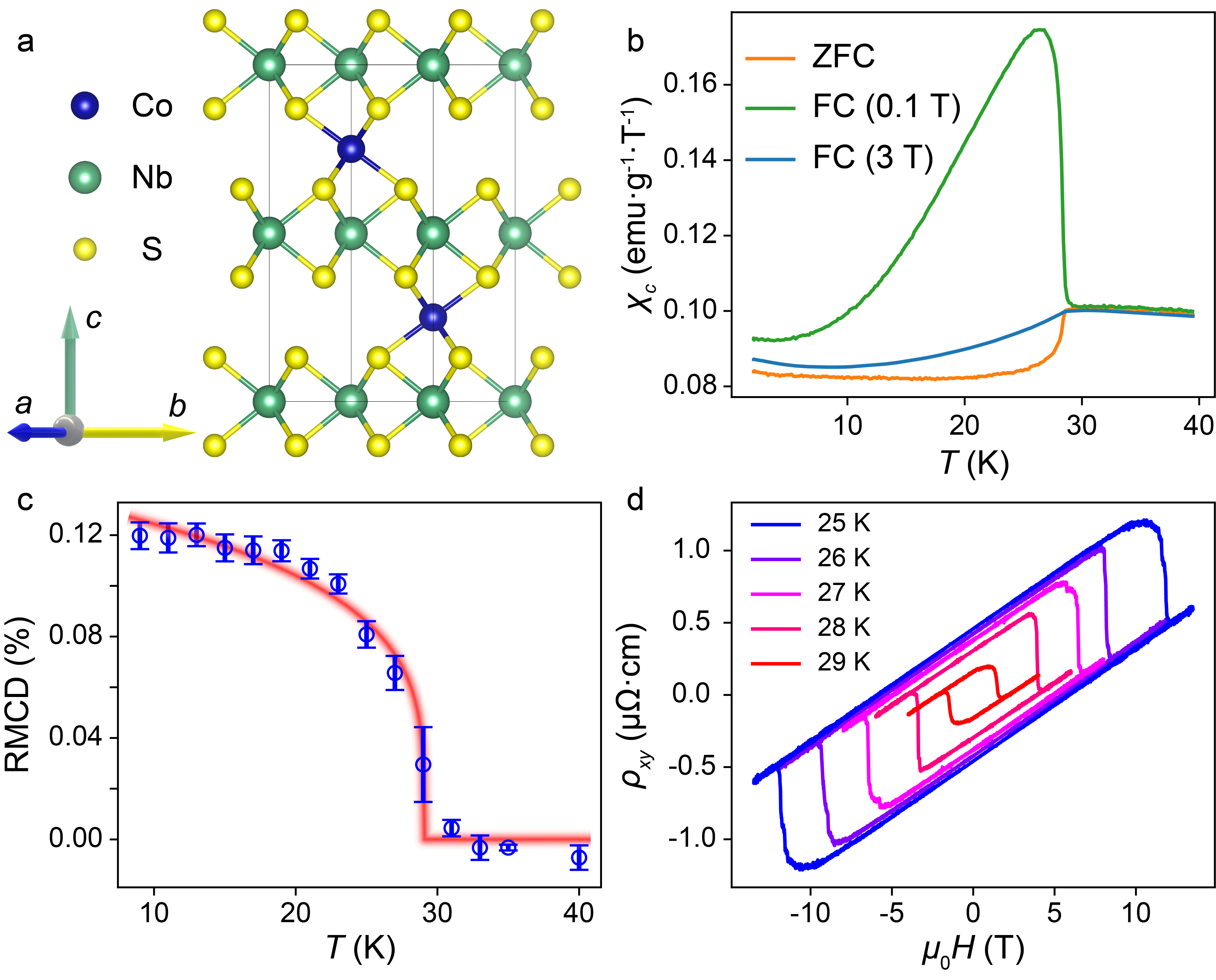}
		\caption{Basic characterizations of $\rm{Co_{1/3}NbS_2}$. (a), The crystal structure of $\rm{Co_{1/3}NbS_2}$ viewed along the axis perpendicular to the $bc$ plane. (b), The measured temperature-dependent out-of-plane susceptibility of single crystal $\rm{Co_{1/3}NbS_2}$ bulk with ZFC, 0.1 T FC, and 3 T FC. (c), RMCD signal \textit{versus} temperature of an exfoliated $\rm{Co_{1/3}NbS_2}$ flake with a thickness of $\sim$130 nm. Red hatching highlights the phase transition, which is used to guide the eye. (d), Transverse resistivity $\rho_{xy}$ \textit{versus} out-of-plane magnetic field of device 1 at different temperatures, where the anomalous and ordinary Hall coefficients can be extracted.}
		\label{Figure1}
	\end{figure}
	
	$\rm{Co_{1/3}NbS_2}$ single crystals were synthesized \textit{via} the chemical vapor transport (CVT) technique (see Methods), and Fig. \ref{Figure1}a shows the crystal structure viewed along the axis perpendicular to the $bc$ plane. The $\rm{Co^{2+}}$ cations ideally reside only at the 2$c$ Wyckoff site, resulting in a $\rm{\sqrt{3}}$-type superstructure lacking inversion symmetry with a space group of $P6_322$. Figure \ref{Figure1}b shows the out-of-plane magnetization of bulk crystals. The apparent inflections at 29 K signify the Néel temperature ($T_N$=29 K), consistent with previous reports\cite{friend1977electrical,ghimire2018large}. The magnetization of $\rm{Co_{1/3}NbS_2}$ below $T_N$ is composed of a tiny rectangular hysteresis loop and a linear canting background (see Supplementary Information Fig. S1). The tiny ferromagnetic component along the $c$-axis results in the abrupt increase of susceptibility at $T_N$ in the 0.1 T field cooling (FC) curve, while the linear background is responsible for nearly the same trend in the zero field cooling (ZFC) and 3 T FC curves (Fig. \ref{Figure1}b). The measured magnetic moment is more than three orders of magnitudes smaller than the spin moment of $\rm{Co^{2+}}$ ions (3.87 $\rm{\mu_B}$)\cite{parkin1983magnetic}, indicating that $\rm{Co_{1/3}NbS_2}$ established a long-range antiferromagnetic ground state below $T_N$ .\par
	
	Considering that another possible order parameter rather than the magnetic moment can break the time-reversal symmetry and induce a non-zero Berry curvature\cite{suzuki2017cluster}, we performed RMCD measurements on the exfoliated $\rm{Co_{1/3}NbS_2}$ flakes. After 6 T FC, the exfoliated flake exhibits a distinct RMCD signal below $T_N$ by $\sim$0.12\% (Fig. \ref{Figure1}d), comparable to most ferromagnetic materials but with a negligible net magnetic moment. In general, the RMCD signal is proportional to the optical transverse conductivity, and thus to the Berry curvature\cite{feng2015large,higo2018large}. Such a large RMCD value optically verifies the large Berry curvature in $\rm{Co_{1/3}NbS_2}$ and may resolve hidden order parameters coupled with Berry curvature. Figure \ref{Figure1}d shows the temperature-dependent Hall measurements of an exfoliated $\rm{Co_{1/3}NbS_2}$ flake with a thickness of $\sim$125 nm, labeled as device 1. A large AHE can be observed, and the critical field $H_c$ increases sharply with decreasing temperature, indicating an extremely large magnetic anisotropy. Both the anomalous and ordinary Hall coefficients can be obtained from the measurements and will be discussed in detail later.\par
	
	\begin{figure}
		\centering
		\includegraphics[width=1.0\linewidth]{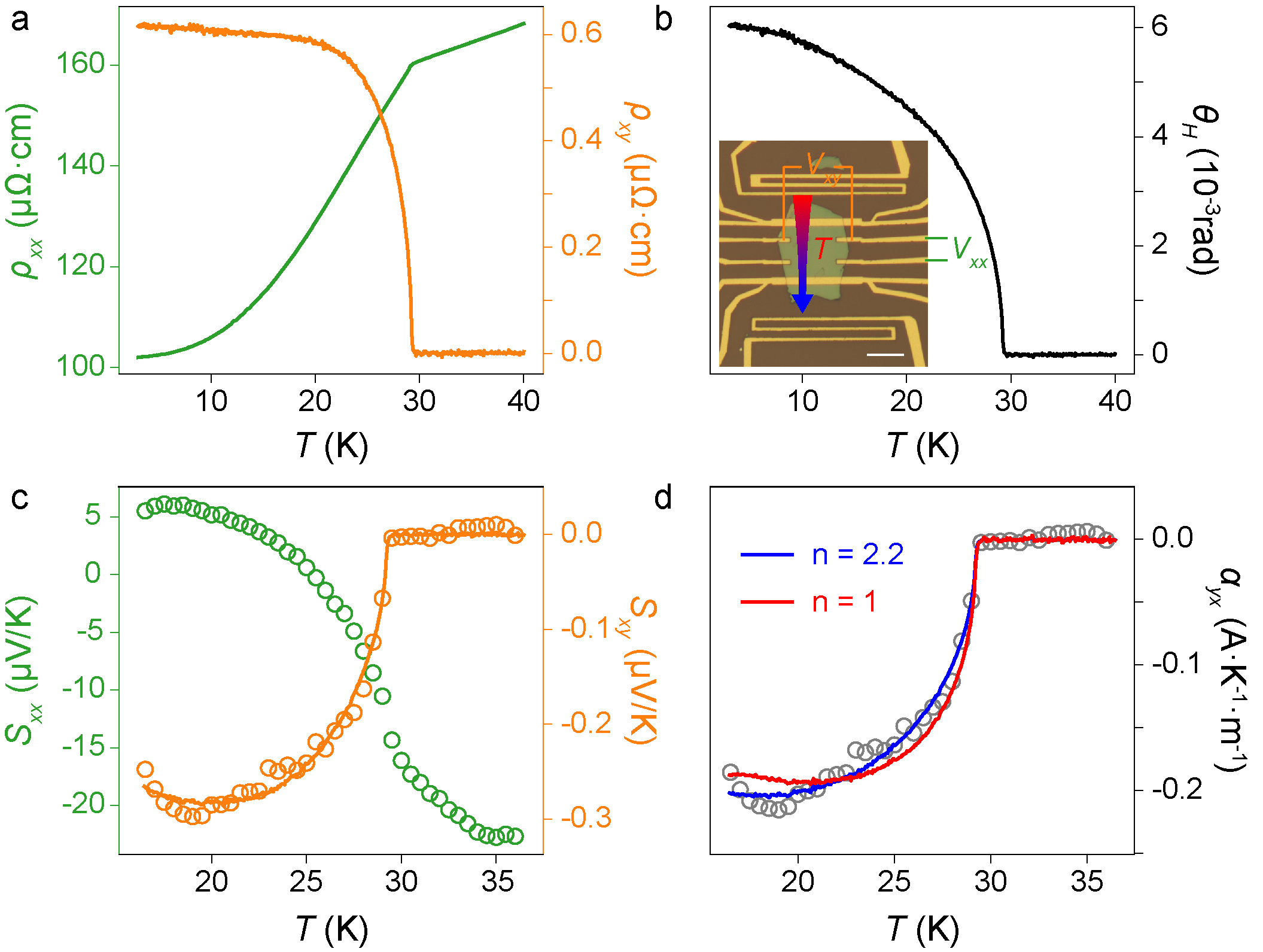}
		\caption{Electrical transport and thermoelectric measurements of $\rm{Co_{1/3}NbS_2}$. (a), Temperature-dependent longitudinal and transverse resistivity. (b), AHE angle $\theta_H=|\frac{\sigma_{xy}}{\sigma_{xx}}|$ \textit{versus} temperature, extracted from (a). The inset shows the optical image of the device used for electrical transport and thermoelectric measurements. The scale bar is 20 $\mu$m. (c), Temperature-dependent Seebeck and Nernst coefficients of $\rm{Co_{1/3}NbS_2}$. Open circles are experimental data, and the solid line represents the best fit of the Mott relation to the Nernst coefficient. The best-fit parameters give $n=2.2$. (d), Extracted thermoelectric conductivity \textit{versus} temperature. The blue curve is the best fit using the Mott relation, and the red curve is the best fit with $n=1$.}
		\label{Figure2}
	\end{figure}
	
	The temperature dependence of the longitudinal and transverse resistivity, $\rho_{xx}$ and $\rho_{xy}$, is shown in Fig. \ref{Figure2}a. $\rho_{xx}$ shows typical metallic behavior, as its value decrease with temperature and drops more sharply after an obvious kink at $T_N$. This may be because electron scattering decreases in the long-range magnetic order, in other words, the relaxation time $\tau$ becomes longer. On the contrary, $\rho_{xy}$ exhibits a sharp increase upon establishing magnetic order and reaches a plateau afterward, showing no signs of decreasing with decreasing temperature. This is incompatible with the decrease of $\rho_{xx}$ with temperature, because in a general sense, the transverse resistivity $\rho_{xy}^{AF}$ scales with $\rho_{xx}$ in a power law\cite{nagaosa2010anomalous}:\par
	\begin{equation}
	\rho_{xy}^{AF}=\lambda M\rho_{xx}^n
	\label{Equation1}
	\end{equation}
	
	\noindent where $M$ is the magnetic moment, or the order parameter that induces Berry curvature, and $\lambda$ is a temperature-independent scaling factor. $n$ is the power constant depending on the AHE mechanism. The skew scattering mechanism, where $\sigma_{xy}^{AF}\sim 1/\tau$, leads to $n=1$, while the intrinsic Berry curvature mechanism, where $\sigma_{xy}^{AF}$ is independent on $\tau$, leads to $n=2$\cite{onoda2008quantum}. Apparently, the resistivity of $\rm{Co_{1/3}NbS_2}$ fails to conform to the positive correlation at low temperatures. The only possibility to explain this discrepancy is that the integrated Berry curvature continues to increase after the formation of the antiferromagnetic order. Consequently, the extracted Hall angle $\theta_H=|\frac{\sigma_{xy}}{\sigma_{xx}}|$ keeps increasing below $T_N$ and reaches 6 mrad at 2 K. Similar behavior is observed in all devices we measured (see Supplementary Information Fig. S2 and Table S1), with a maximum Hall angle of $\sim$0.014, comparable to other 2D ferromagnetic materials\cite{kim2018large,chen2021anomalous}.\par
	
	To elucidate the scaling behavior between $\rho_{xx}$ and $\rho_{xy}$ and the origin of AHE, we measured the thermoelectric coefficients of $\rm{Co_{1/3}NbS_2}$. Distinct from charge flow transport, thermoelectric signals detect the flow of entropy and thus serve as sensitive probes of the electronic properties of Fermi surfaces, especially the Berry curvature\cite{xiao2006berry,wuttke2019berry}. The inset of Fig. \ref{Figure2}b shows the optical image of device 1. To generate a lateral temperature gradient $\nabla T$, a heater at one side of the device is heated by driving an a.c. current. The actual temperature and temperature gradient are calibrated by four-probe resistance measurements of the temperature sensors placed on both sides of the sample, and the Seebeck and Nernst signals are obtained by measuring the $2\omega$ voltage signals of the source-drain and Hall electrodes of the sample. The detailed process of temperature calibration is discussed in Supplementary Information Fig. S3 and S4. As shown in Fig. \ref{Figure2}c, the Nernst coefficient $S_{xy}$ becomes non-zero below $T_N$ and reaches as large as $\rm{0.3\ \mu V/K}$. The Seebeck coefficient, $S_{xx}$, is negative at high temperatures and becomes positive below $\sim$24 K, in good accordance with the measurements for bulk crystals\cite{barivsic2011high} and reproducible in all samples we measured (Supplementary Information Fig. S2).\par
	
	We note that the sign change of $S_{xx}$ should not be attributed to any phonon-drag effect, since the temperature is much lower than the expected value of $\theta_D/5$\cite{barivsic2011high,inoue1986specific,popvcevic2020electronic} ($\theta_D$ is the Debye temperature). Neither the $\rho_{xy}$ or $S_{xy}$ signals exhibit any abnormal changes around 24 K, ruling out the possibility of a complete change in carrier type. Actually, we can obtain the effective carrier concentration by a linear fit of the ordinary Hall effect extracted from Fig. \ref{Figure1}d (see Supplementary Information Fig. S5). The ordinary Hall effect indicates that the material is \textit{p}-type conduction and the hole concentration decreases with decreasing temperature, which is consistent with previous reports\cite{ghimire2018large,tenasini2020giant}, but incompatible with the trend of $S_{xx}$ turning from negative to positive. These observations can lead to the conclusion of the coexistence of hole and electron carriers in this material. The ordinary Hall effect is mainly contributed by hole carrier, which is the majority carrier provided by the $\rm{NbS_2}$ bands\cite{tanaka2022large}. In contrast, the $S_{xx}$ signal is dominated by the electron band near the Fermi level, which is resulted from the intercalated $\rm{Co^{2+}}$ cations. $S_{xx}$ should approach zero at low temperatures as the entropy should vanish at $T\rightarrow 0$, but the contribution of holes gradually constitutes a larger proportion and exceeds that of electrons, resulting in the sign change from negative to positive. Consequently, the sign change of $S_{xx}$ may indicate a smooth electronic transition or a Fermi level shift, which is also manifested in the temperature-dependent carrier concentration (Supplementary Information Fig. S5). The overall hole concentration kinks between 20 K and 25 K and stops decreasing at lower temperatures, confirming our above conjecture. Differently, since the absence of sign change in $\rho_{xy}$ and $S_{xy}$, we can conclude that only one type energy band, presumably the electron band, is non-trivial and contributes to AHE and anomalous Nernst effect (ANE). The sign change of $S_{xx}$ and finite $S_{xy}$ naturally lead to a divergent Nernst angle by $\theta_N = \frac{S_{yx}}{S_{xx}}$. As a comparison, the maximum Nernst angle at the lowest measurement temperature of the devices reaches 0.12, and the ratio between anomalous Nernst coefficient to the spontaneous magnetization $S_{yx}/M$ reaches $\rm{10^3\ (\mu V \cdot K^{-1}\cdot (\mu_B f.u.^{-1})^{-1})}$ (see Supplementary Information Table S1). These values are much larger than the common FM metals\cite{xu2019large,hasegawa2015material,chuang2017enhancement} and comparable to other topological materials\cite{sakai2018giant,yang2020giant,pan2022giant}.\par
	
	In addition to electrical transport, thermoelectric measurements provide us with another degree of freedom to explore the underlying mechanism of AHE. By definition, the Seebeck coefficient is related to other transport parameters by:\par
	
	\begin{equation}
	S_{y x}=\frac{1}{\sigma_{x x}}\left(\alpha_{y x}-\sigma_{y x} S_{x x}\right)
	\label{Equation2}
	\end{equation}
	
	\noindent with the measured electrical conductivity and thermoelectric coefficient, we can obtain the transverse thermoelectric conductivity $\alpha_{xy}$ by the above equation, as shown in Fig. \ref{Figure2}d. Furthermore, the Mott relation describes the relationship between $\alpha_{xy}$ and $\sigma_{xy}$ by $\alpha_{xy}=-\frac{\pi^2 k_{B}^2T}{3 e} \left(\frac{\partial \sigma_{xy}}{\partial \varepsilon}\right)\bigg|_{\varepsilon_{F}}$. Substituting Eq. \ref{Equation1} into the Mott relation and combining Eq. \ref{Equation2}, we can eliminate the common factor (order parameter $M$) and obtain the modified Mott relation containing only four transport parameters\cite{pu2008mott}:
	
	\begin{equation}
	S_{y x}=\frac{\rho_{x y}}{\rho_{x x}}\left(T \frac{\pi^2 k_B^2}{3 e} \frac{\lambda^{\prime}}{\lambda}-(n-1) S_{x x}\right)
	\label{Equation3}
	\end{equation}
	
	\noindent and
	
	\begin{equation}
	\alpha_{y x}=\frac{\rho_{x y}}{\rho_{x x}^2}\left(T \frac{\pi^2 k_B^2}{3 e} \frac{\lambda^{\prime}}{\lambda}-(n-2) S_{x x}\right)
	\label{Equation4}
	\end{equation}
	
	\noindent where $\lambda$ is the same parameter as in Eq. \ref{Equation1}, and $\lambda^{\prime}$ is the energy derivation of $\lambda$. Both of them should be constants. These two equations verify the Mott relation and determine the exponent $n$ without including the unknown temperature-dependent order parameter. The best-fit parameters give $n=2.2$ (Fig. \ref{Figure2}c and d), which is close to $n=2$ rather than $n=1$ (deviates significantly from the experimental data as shown by the red curve in Fig. \ref{Figure2}d). This $n$ value indicates that the anomalous Hall resistivity in $\rm{Co_{1/3}NbS_2}$ is independent of the relaxation time $\tau$ and thus likely arises from the intrinsic Berry curvature, as suggested by the previous theories\cite{park2022first,zhang2023chiral,lu2022understanding,takagi2023spontaneous}. More importantly, the fitting results verify that Eq. \ref{Equation1} remains valid in our samples, but there is a temperature-dependent term that causes the Berry curvature to increase at low temperatures, so although $\rho_{xx}$ decreases, $\rho_{xy}$ continues to increase. This term could arise from the large thermal fluctuations of the frustrated magnetic order, or from the aforementioned electronic transitions or Fermi level shifts. Generally, large magnetic anisotropy can lead to a rapid stabilization of the order parameter below $T_N$, so the latter explanation seems more plausible in this system.\par
	
	\begin{figure}
		\centering
		\renewcommand{\figurename}{Figure}
		\includegraphics[width=1.0\linewidth]{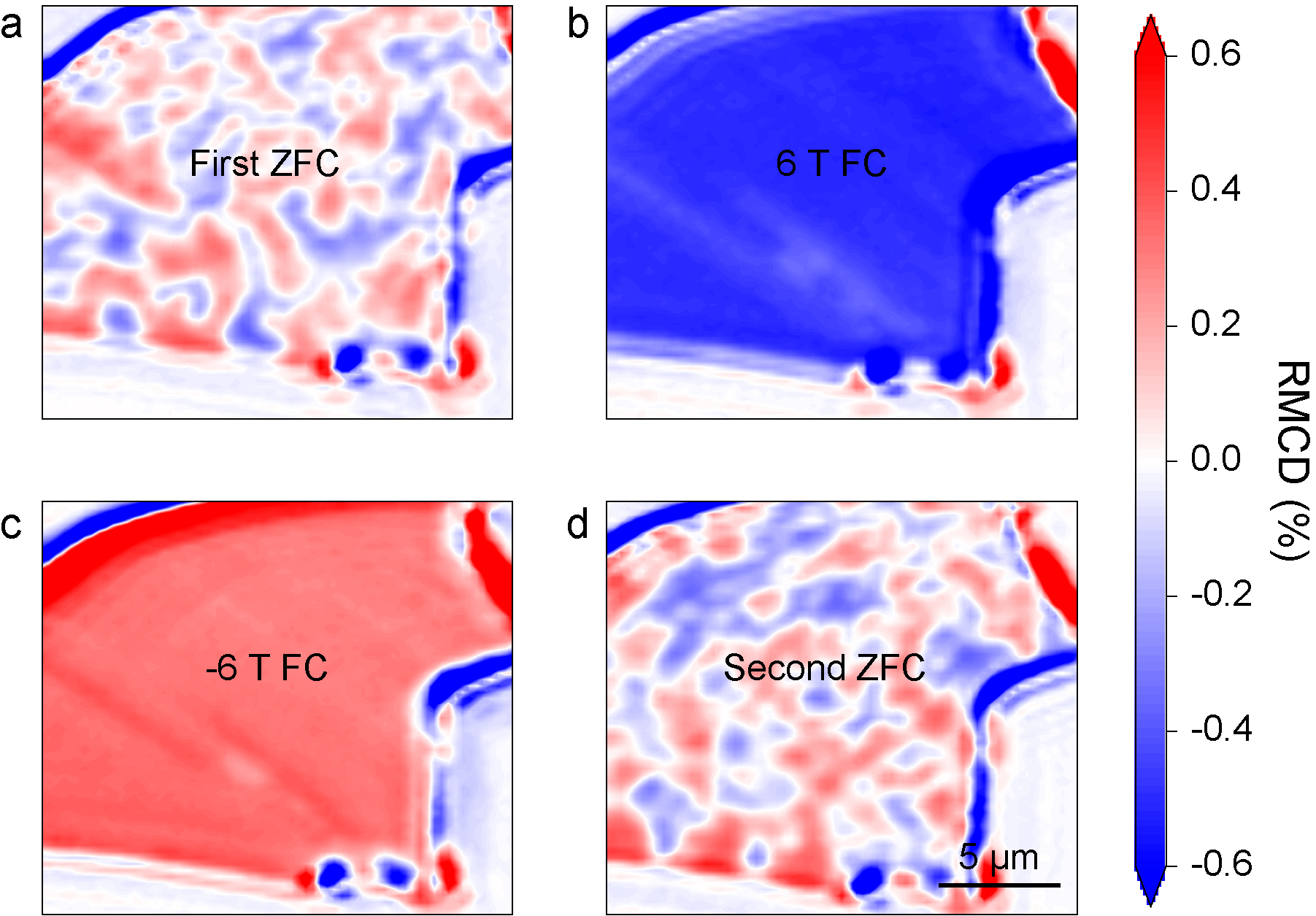}
		\caption{RMCD mapping of exfoliated 100 nm thick $\rm{Co_{1/3}NbS_2}$ flake at 2 K. (a), RMCD mapping after the first ZFC. (b, c), RMCD mappings after 6 T (b) and $-$6 T (c) FC. (d), RMCD mapping after the second ZFC. }
		\label{Figure3}
	\end{figure}
	
	After understanding the underlying mechanism of the anomalous transport, we turn to magnetic domains and magnetic reversal behavior in $\rm{Co_{1/3}NbS_2}$. We have demonstrated in Fig. \ref{Figure1}c that the order parameter in $\rm{Co_{1/3}NbS_2}$ coupled with Berry curvature can be resolved by the RMCD signal. Therefore, by scanning the entire flake with a laser spot ($\sim$1.5 $\mu$m), we can obtain the spatially resolved RMCD signals and therefore detect the domain structure of the exfoliated samples, as shown in Fig. \ref{Figure3}. Optical and atomic force microscopy height images of the same exfoliated flake, together with the single-point RMCD signal \textit{versus} temperature after $\pm6\ \rm{T}$ FC can be seen in Supplementary Information Fig. S6. The main observations are summarized below. First, after ZFC (Fig. \ref{Figure3}a), the RMCD signal exhibits clear spatial variation between $\pm0.5\%$, in accordance with the RMCD value after FC (see Supplementary Information Fig. S6). This mapping unambiguously demonstrates a multi-domain structure with a domain size of several $\mu$m, which is also manifested in the initial magnetization in AHE measurements (see Supplementary Information Fig. S7). After $6\ \rm{T}$ or $-6\ \rm{T}$ field cooling, the whole sample exhibits a homogeneous but opposite RMCD signal (Fig. \ref{Figure3}b-c), showing a single domain structure. The extremely large values at the edges of the sample are artifacts due to the protrusion of the exfoliated sample and do not contain useful information, as indicated in Supplementary Information Fig. S7b. We then performed another ZFC measurement and the domains were completely redistributed (Fig. \ref{Figure3}d), implying that the domain structure was randomly formed upon each cooling from a higher temperature, rather than determined by a series of pinning sites. Line cuts from the four mappings are shown in Supplementary Information Fig. S6. The domains after different cooling processes exhibit similar RMCD values, confirming the reliability of our measurements.\par

    The domains observed in $\rm{Co_{1/3}NbS_2}$ may arise from the spin chirality of spatially distributed all-in-all-out domains as described in Ref. \cite{takagi2023spontaneous}, or simply caused by domains with slightly canted spins towards the $c$-axi of a collinear structure as described in Ref. \cite{lu2022understanding}. If the second scenario applies, there should also be three types of domains with symmetry-related in-plane $q$ vectors and equal weights\cite{parkin1983magnetic,tenasini2020giant}, but unfortunately, this cannot be resolved by RMCD measurements. Nonetheless, direct observation of magnetic domains in micro-sized samples is a crucial step towards correctly determining spin configurations in $\rm{Co_{1/3}NbS_2}$, since the magnetic symmetry in individual domains can be further probed by second-harmonic generation signals\cite{sun2019giant,ni2021direct} or other sensitive techniques.\par
	
	\begin{figure}
		\centering
		\renewcommand{\figurename}{Figure}
		\includegraphics[width=1.0\linewidth]{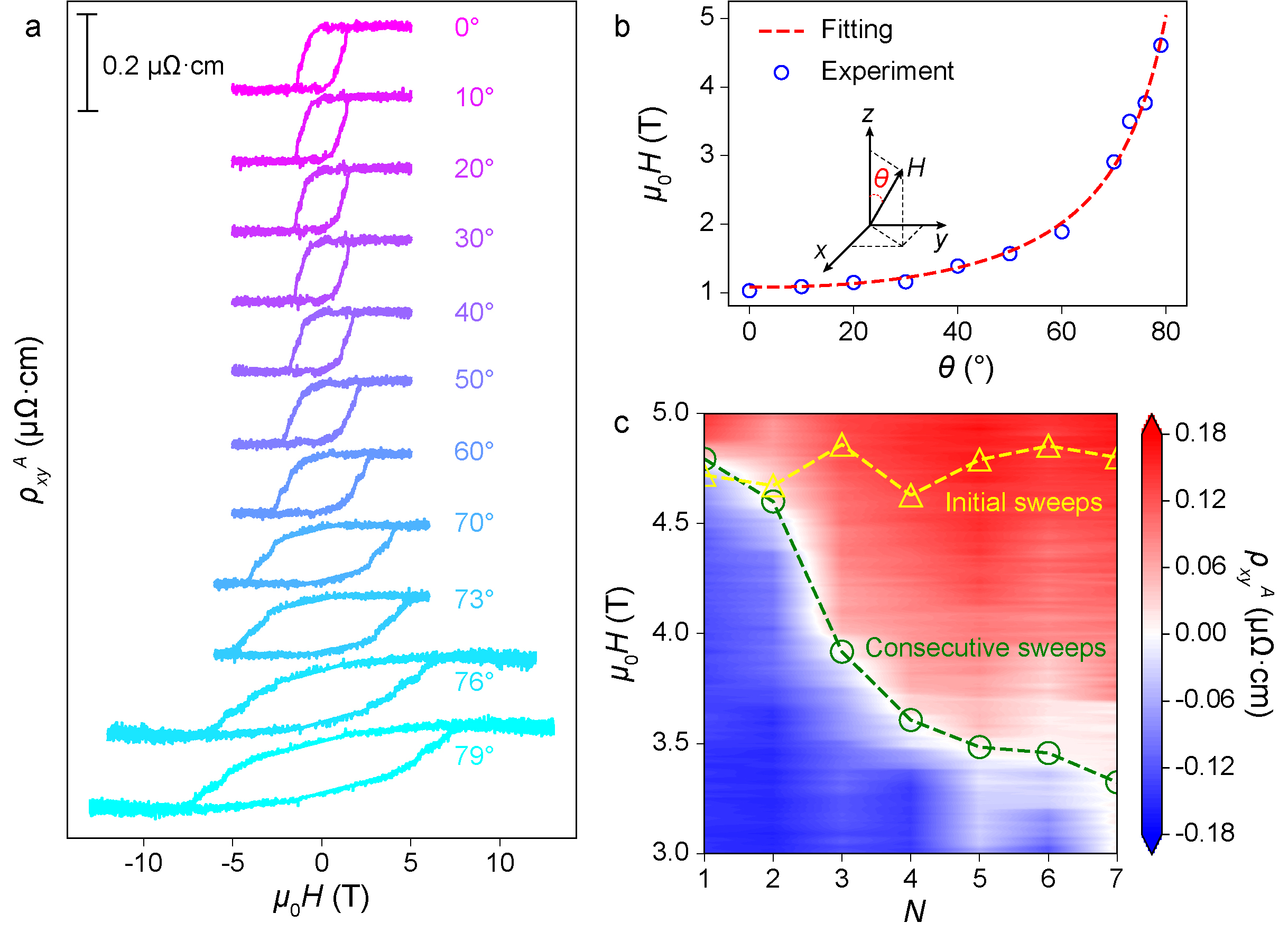}
		\caption{Transport behavior determined by domain motion. (a), Anomalous resistivity $\rho_{xy}^A$ \textit{versus} $\mu_{\rm{0}}H$ under different field angles measured at 29 K. The sample lies in the $x-y$ plane, and the magnetic field $H$ is rotated from the $z$ axis to an in-plane axis. $\theta$ is the angle between $H$ and the $z$ axis. Each curve is shifted by 0.15 $\rm{\mu\Omega\cdot cm}$ relative to the curve below. (b), Coercive field $H_c$ as a function of $\theta$. $H_c$ is defined as the intercept of $\rho_{xy}^A$ on the $H$ axis extracted from (a). The dashed red line represents the fit of the Kondorsky model. (c), $\rho_{xy}^A$ during several consecutive magnetic field sweeps at 27.5 K. $H_c$ decreases as the number of sweeps increases. The green dashed line plots the decreasing $H_c$ \textit{versus} the number of sweeps $N$, while the yellow dashed line plots the $H_c$ extracted from seven initial sweeps after cooling down from 50 K. The data were obtained from a 120 nm thick $\rm{Co_{1/3}NbS_2}$ device, labeled as device 2.}
		\label{Figure4}
	\end{figure}
	
	Additionally, the multi-domain structure also determines the magnetic reversal process in $\rm{Co_{1/3}NbS_2}$. To determine the magnetic reversal model, we measured the AHE hysteresis loops at different magnetic field angles, as shown in Fig. \ref{Figure4}a. When the field is rotated from the $z$ axis to an in-plane axis, the coercive field $H_c$ increases monotonically and can be well fitted by the $1/\rm{cos}\theta$ function (Fig. \ref{Figure4}b), while the AHE value remains almost unchanged. The above phenomena indicate that the magnetic reversal in $\rm{Co_{1/3}NbS_2}$ follows the Kondorsky model\cite{schumacher1991modification} rather than the coherent Stoner–Wohlfarth model\cite{stoner1948mechanism}. The energy of domain wall propagation is much lower than the magnetic anisotropy energy, so $H_c$ is determined by the competition between Zeemann energy and domain wall energy. These are consistent with the observed sharp increase of $H_c$ with decreasing temperature (Fig. \ref{Figure1}d) and the multi-domain structure (Fig. \ref{Figure3}). Moreover, the AHE reversal appears to be a slow ramping with the increasing external field instead of a steep step, and the slope becomes lower when the field points to the in-plane axis, implying that the domain wall motion is almost steady and slow.\par
	
	The domain-determined magnetic reversal also leads to another manifestation, that is, the decrease of $H_c$ with consecutive sweeps of the magnetic field. Figure \ref{Figure4}c shows the 2D plot of $\rho_{xy}^A$ as a function of magnetic field and sweeping number $N$ at 27.5 K. The green dashed line illustrates the extracted $H_c$ \textit{versus} $N$, which drops sharply over the first few sweeps and gradually becomes steady. Seven consecutive sweeps reduce $H_c$ in device 2 by 28\%. However, $H_c$ can be set back to the initial value by raising the temperature above $T_N$ and cooling back to 27.5 K, which we call the initialization process. We raised the temperature to 50 K and then cooled it back, and then directly carried out the AHE measurement. The initialization process was repeated seven times, and the seven extracted $H_c$ are plotted in the yellow dashed line in Fig. \ref{Figure4}c, exhibiting consistent values with perfect reproducibility. Naturally, after initialization, $H_c$ will follow the same trend of decreasing if we perform consecutive magnetic sweeps. In other words, the hysteresis loop can be manipulated by continuously sweeping the external field, and its $H_c$ information will be memorized but can be erased after the temperature rises above $T_N$ and cools down again. It is important to emphasize that this memory effect didn't appear by chance in a single device, but was observed in all the devices we measured. We also repeated the same measurements in device 1 (see Supplementary Information Fig. S8), and the results were in good agreement with device 2, indicating that the manipulation of $H_c$ is an intrinsic property in exfoliated $\rm{Co_{1/3}NbS_2}$ flakes. Based on the established knowledge that the domain structure redistributes after each cooling (Fig. \ref{Figure3}a and d), and the value of $H_c$ is determined by the energy of domain wall motion (Fig. \ref{Figure4}a-b), we can infer that the decrease in $H_c$ upon consecutive sweeps is not a trivial consequence of the change in pinning site, but rather a modification in the domain wall propagation energy. Actually, the manipulation of $H_c$ is likely to originate from the memory effect of chiral domain walls\cite{li2019chiral} or the intrinsic exchange bias effect\cite{maniv2021exchange} observed in similar systems, which requires confirmation by further experiments and theoretical calculations, and may provide a basis for the manipulation of non-collinear antiferromagnets.
	
	\section{Conclusion}
	
	In summary, we investigated the large AHE, ANE, and the magnetic domain-related behavior in exfoliated $\rm{Co_{1/3}NbS_2}$ flakes. The Mott relation between transport and thermoelectric coefficients is verified, and the fitting results unambiguously indicate an intrinsic large Berry curvature. A series of phenomena imply large magnetic fluctuations or electronic transitions at low temperatures. Furthermore, we observed a large RMCD signal and resolved magnetic domains by RMCD mapping, optically evidencing the large Berry curvature. Combined with transport measurements, we concluded that the magnetization reversal in $\rm{Co_{1/3}NbS_2}$ is dominated by the domain wall motion. The domain wall energy determined $H_c$ decreases monotonically with consecutive field sweeps, which probably indicates a memory effect of domain walls. Our work probes the intrinsic Berry curvature and provides an explicit picture of AHE and magnetic reversal mechanism in $\rm{Co_{1/3}NbS_2}$ without including its controversial magnetic order. A detailed understanding of this antiferromagnet will provide useful information for future spin caloritronics applications.
 
    \section{Methods}
	\noindent
    \textbf{Crystal growth and magnetic characterizations}.
    High-quality $\rm{Co_{1/3}NbS_2}$ single crystals were grown by the chemical vapor transport method. Co, Nb, and S powders with a ratio of 1:3:6 were sealed in a quartz tube, then put into a furnace, heated to 800 °C, and kept for 5 days to prepare polycrystalline precursor. The resulting powders (1 g) and transport agent iodine (15 $\rm{mg/cm^3}$) were then sealed in a quartz tube to grow single crystals with a temperature gradient set between 950 °C (source) and 850 °C (products) for 10 days. Finally, hexagonal plate-shaped single crystals were obtained, which are easy to exfoliate. The thickness of the ultrathin samples was verified by the atomic force microscopy characterization using an Oxford Cypher S system in tapping mode. Magnetization measurements were performed by standard modules of a Quantum Design PPMS.
    \bigskip

    \noindent
    \textbf{Electrical and thermoelectric measurements}. Metal contact electrodes of Cr/Au (10/80 nm) were defined using electron beam lithography, electron beam evaporation, and lift-off processes on the exfoliated flakes. The devices were then loaded into a physical property measurement system (Cryomagnetics) with a magnetic field up to 14 T. A.c. voltage measurements were performed with Stanford Research SR830 lock-in amplifiers using the standard four-point method. $\rho_{xx}$ and $\rho_{xy}$ were measured under an a.c. current of 10 $\rm{\mu A}$ at 17.777 Hz, while $\S_{xx}$ and $\S_{xy}$ were measured under an a.c. heater current of 0.2 mA at 3.777 Hz. Temperature perturbation was calibrated by four-probe resistances of the thermocouples. Details of thermoelectric measurements and temperature calibration are presented in Supplementary Information Fig. S3-4 and the following discussions.
    \bigskip

    \noindent
    \textbf{RMCD measurements}. The RMCD measurements were performed based on the Attocube closed-cycle cryostat (attoDRY2100) down to 1.6 K and up to 9 T in the out-of-plane direction. The linearly polarized light of the 633 nm HeNe laser was modulated between left and right circular polarization by a photoelastic modulator (PEM) and focused onto the sample through a high numerical aperture (0.82) objective. The reflected light was detected by a photomultiplier tube (THORLABS PMT1001/M). The magnetic reversal under the external magnetic field was detected by the RMCD signal determined by the ratio of the a.c. component of PEM at 50.052 kHz and the a.c. component of the chopper at 779 Hz (detected by a two-channel lock-in amplifier Zurich HF2LI). RMCD mapping was implemented by moving the piezo sample stage.
	
	\begin{acknowledgments}
		This work was supported by the National Natural Science Foundation of China (No. 12241401 and No. 12250007), the National Key R\&D Program of China (Grants No. 2022YFA1203902 and No. 2018YFA0306900), and Beijing Natural Science Foundation (Grant No. JQ21018). T. X. acknowledges support from the National Key R\&D Program of China (Grant No. 2019YFA0308602), and the National Natural Science Foundation of China (Grant Nos. 12074425 and 11874422).

	\end{acknowledgments}
	
	\bibliography{Reference}

\begin{thebibliography}{52}%
\makeatletter
\providecommand \@ifxundefined [1]{%
 \@ifx{#1\undefined}
}%
\providecommand \@ifnum [1]{%
 \ifnum #1\expandafter \@firstoftwo
 \else \expandafter \@secondoftwo
 \fi
}%
\providecommand \@ifx [1]{%
 \ifx #1\expandafter \@firstoftwo
 \else \expandafter \@secondoftwo
 \fi
}%
\providecommand \natexlab [1]{#1}%
\providecommand \enquote  [1]{``#1''}%
\providecommand \bibnamefont  [1]{#1}%
\providecommand \bibfnamefont [1]{#1}%
\providecommand \citenamefont [1]{#1}%
\providecommand \href@noop [0]{\@secondoftwo}%
\providecommand \href [0]{\begingroup \@sanitize@url \@href}%
\providecommand \@href[1]{\@@startlink{#1}\@@href}%
\providecommand \@@href[1]{\endgroup#1\@@endlink}%
\providecommand \@sanitize@url [0]{\catcode `\\12\catcode `\$12\catcode
  `\&12\catcode `\#12\catcode `\^12\catcode `\_12\catcode `\%12\relax}%
\providecommand \@@startlink[1]{}%
\providecommand \@@endlink[0]{}%
\providecommand \url  [0]{\begingroup\@sanitize@url \@url }%
\providecommand \@url [1]{\endgroup\@href {#1}{\urlprefix }}%
\providecommand \urlprefix  [0]{URL }%
\providecommand \Eprint [0]{\href }%
\providecommand \doibase [0]{https://doi.org/}%
\providecommand \selectlanguage [0]{\@gobble}%
\providecommand \bibinfo  [0]{\@secondoftwo}%
\providecommand \bibfield  [0]{\@secondoftwo}%
\providecommand \translation [1]{[#1]}%
\providecommand \BibitemOpen [0]{}%
\providecommand \bibitemStop [0]{}%
\providecommand \bibitemNoStop [0]{.\EOS\space}%
\providecommand \EOS [0]{\spacefactor3000\relax}%
\providecommand \BibitemShut  [1]{\csname bibitem#1\endcsname}%
\let\auto@bib@innerbib\@empty
\bibitem [{\citenamefont {Nagaosa}\ \emph {et~al.}(2010)\citenamefont
  {Nagaosa}, \citenamefont {Sinova}, \citenamefont {Onoda}, \citenamefont
  {MacDonald},\ and\ \citenamefont {Ong}}]{nagaosa2010anomalous}%
  \BibitemOpen
  \bibfield  {author} {\bibinfo {author} {\bibfnamefont {N.}~\bibnamefont
  {Nagaosa}}, \bibinfo {author} {\bibfnamefont {J.}~\bibnamefont {Sinova}},
  \bibinfo {author} {\bibfnamefont {S.}~\bibnamefont {Onoda}}, \bibinfo
  {author} {\bibfnamefont {A.~H.}\ \bibnamefont {MacDonald}},\ and\ \bibinfo
  {author} {\bibfnamefont {N.~P.}\ \bibnamefont {Ong}},\ }\bibfield  {title}
  {\bibinfo {title} {{Anomalous Hall effect}},\ }\href@noop {} {\bibfield
  {journal} {\bibinfo  {journal} {Rev. Mod. Phys.}\ }\textbf {\bibinfo {volume}
  {82}},\ \bibinfo {pages} {1539} (\bibinfo {year} {2010})}\BibitemShut
  {NoStop}%
\bibitem [{\citenamefont {Chen}\ \emph {et~al.}(2014)\citenamefont {Chen},
  \citenamefont {Niu},\ and\ \citenamefont {MacDonald}}]{chen2014anomalous}%
  \BibitemOpen
  \bibfield  {author} {\bibinfo {author} {\bibfnamefont {H.}~\bibnamefont
  {Chen}}, \bibinfo {author} {\bibfnamefont {Q.}~\bibnamefont {Niu}},\ and\
  \bibinfo {author} {\bibfnamefont {A.~H.}\ \bibnamefont {MacDonald}},\
  }\bibfield  {title} {\bibinfo {title} {{Anomalous Hall effect arising from
  noncollinear antiferromagnetism}},\ }\href@noop {} {\bibfield  {journal}
  {\bibinfo  {journal} {Phys. Rev. Lett.}\ }\textbf {\bibinfo {volume} {112}},\
  \bibinfo {pages} {017205} (\bibinfo {year} {2014})}\BibitemShut {NoStop}%
\bibitem [{\citenamefont {Martin}\ and\ \citenamefont
  {Batista}(2008)}]{martin2008itinerant}%
  \BibitemOpen
  \bibfield  {author} {\bibinfo {author} {\bibfnamefont {I.}~\bibnamefont
  {Martin}}\ and\ \bibinfo {author} {\bibfnamefont {C.}~\bibnamefont
  {Batista}},\ }\bibfield  {title} {\bibinfo {title} {Itinerant electron-driven
  chiral magnetic ordering and spontaneous quantum hall effect in triangular
  lattice models},\ }\href@noop {} {\bibfield  {journal} {\bibinfo  {journal}
  {Phys. Rev. Lett.}\ }\textbf {\bibinfo {volume} {101}},\ \bibinfo {pages}
  {156402} (\bibinfo {year} {2008})}\BibitemShut {NoStop}%
\bibitem [{\citenamefont {Shindou}\ and\ \citenamefont
  {Nagaosa}(2001)}]{shindou2001orbital}%
  \BibitemOpen
  \bibfield  {author} {\bibinfo {author} {\bibfnamefont {R.}~\bibnamefont
  {Shindou}}\ and\ \bibinfo {author} {\bibfnamefont {N.}~\bibnamefont
  {Nagaosa}},\ }\bibfield  {title} {\bibinfo {title} {{Orbital ferromagnetism
  and anomalous Hall effect in antiferromagnets on the distorted fcc
  lattice}},\ }\href@noop {} {\bibfield  {journal} {\bibinfo  {journal} {Phys.
  Rev. Lett.}\ }\textbf {\bibinfo {volume} {87}},\ \bibinfo {pages} {116801}
  (\bibinfo {year} {2001})}\BibitemShut {NoStop}%
\bibitem [{\citenamefont {Taguchi}\ \emph {et~al.}(2001)\citenamefont
  {Taguchi}, \citenamefont {Oohara}, \citenamefont {Yoshizawa}, \citenamefont
  {Nagaosa},\ and\ \citenamefont {Tokura}}]{taguchi2001spin}%
  \BibitemOpen
  \bibfield  {author} {\bibinfo {author} {\bibfnamefont {Y.}~\bibnamefont
  {Taguchi}}, \bibinfo {author} {\bibfnamefont {Y.}~\bibnamefont {Oohara}},
  \bibinfo {author} {\bibfnamefont {H.}~\bibnamefont {Yoshizawa}}, \bibinfo
  {author} {\bibfnamefont {N.}~\bibnamefont {Nagaosa}},\ and\ \bibinfo {author}
  {\bibfnamefont {Y.}~\bibnamefont {Tokura}},\ }\bibfield  {title} {\bibinfo
  {title} {{Spin chirality, Berry phase, and anomalous Hall effect in a
  frustrated ferromagnet}},\ }\href@noop {} {\bibfield  {journal} {\bibinfo
  {journal} {Science}\ }\textbf {\bibinfo {volume} {291}},\ \bibinfo {pages}
  {2573} (\bibinfo {year} {2001})}\BibitemShut {NoStop}%
\bibitem [{\citenamefont {Ohgushi}\ \emph {et~al.}(2000)\citenamefont
  {Ohgushi}, \citenamefont {Murakami},\ and\ \citenamefont
  {Nagaosa}}]{ohgushi2000spin}%
  \BibitemOpen
  \bibfield  {author} {\bibinfo {author} {\bibfnamefont {K.}~\bibnamefont
  {Ohgushi}}, \bibinfo {author} {\bibfnamefont {S.}~\bibnamefont {Murakami}},\
  and\ \bibinfo {author} {\bibfnamefont {N.}~\bibnamefont {Nagaosa}},\
  }\bibfield  {title} {\bibinfo {title} {{Spin anisotropy and quantum Hall
  effect in the kagom{\'e} lattice: Chiral spin state based on a
  ferromagnet}},\ }\href@noop {} {\bibfield  {journal} {\bibinfo  {journal}
  {Phys. Rev. B}\ }\textbf {\bibinfo {volume} {62}},\ \bibinfo {pages} {R6065}
  (\bibinfo {year} {2000})}\BibitemShut {NoStop}%
\bibitem [{\citenamefont {Nakatsuji}\ \emph {et~al.}(2015)\citenamefont
  {Nakatsuji}, \citenamefont {Kiyohara},\ and\ \citenamefont
  {Higo}}]{nakatsuji2015large}%
  \BibitemOpen
  \bibfield  {author} {\bibinfo {author} {\bibfnamefont {S.}~\bibnamefont
  {Nakatsuji}}, \bibinfo {author} {\bibfnamefont {N.}~\bibnamefont
  {Kiyohara}},\ and\ \bibinfo {author} {\bibfnamefont {T.}~\bibnamefont
  {Higo}},\ }\bibfield  {title} {\bibinfo {title} {{Large anomalous Hall effect
  in a non-collinear antiferromagnet at room temperature}},\ }\href@noop {}
  {\bibfield  {journal} {\bibinfo  {journal} {Nature}\ }\textbf {\bibinfo
  {volume} {527}},\ \bibinfo {pages} {212} (\bibinfo {year}
  {2015})}\BibitemShut {NoStop}%
\bibitem [{\citenamefont {Nayak}\ \emph {et~al.}(2016)\citenamefont {Nayak},
  \citenamefont {Fischer}, \citenamefont {Sun}, \citenamefont {Yan},
  \citenamefont {Karel}, \citenamefont {Komarek}, \citenamefont {Shekhar},
  \citenamefont {Kumar}, \citenamefont {Schnelle}, \citenamefont {K{\"u}bler}
  \emph {et~al.}}]{nayak2016large}%
  \BibitemOpen
  \bibfield  {author} {\bibinfo {author} {\bibfnamefont {A.~K.}\ \bibnamefont
  {Nayak}}, \bibinfo {author} {\bibfnamefont {J.~E.}\ \bibnamefont {Fischer}},
  \bibinfo {author} {\bibfnamefont {Y.}~\bibnamefont {Sun}}, \bibinfo {author}
  {\bibfnamefont {B.}~\bibnamefont {Yan}}, \bibinfo {author} {\bibfnamefont
  {J.}~\bibnamefont {Karel}}, \bibinfo {author} {\bibfnamefont {A.~C.}\
  \bibnamefont {Komarek}}, \bibinfo {author} {\bibfnamefont {C.}~\bibnamefont
  {Shekhar}}, \bibinfo {author} {\bibfnamefont {N.}~\bibnamefont {Kumar}},
  \bibinfo {author} {\bibfnamefont {W.}~\bibnamefont {Schnelle}}, \bibinfo
  {author} {\bibfnamefont {J.}~\bibnamefont {K{\"u}bler}}, \emph {et~al.},\
  }\bibfield  {title} {\bibinfo {title} {{Large anomalous Hall effect driven by
  a nonvanishing Berry curvature in the noncolinear antiferromagnet
  $\rm{Mn_3Ge}$}},\ }\href@noop {} {\bibfield  {journal} {\bibinfo  {journal}
  {Sci. Adv.}\ }\textbf {\bibinfo {volume} {2}},\ \bibinfo {pages} {e1501870}
  (\bibinfo {year} {2016})}\BibitemShut {NoStop}%
\bibitem [{\citenamefont {Liu}\ \emph {et~al.}(2017)\citenamefont {Liu},
  \citenamefont {Zhang}, \citenamefont {Liu}, \citenamefont {Ding},
  \citenamefont {Liu}, \citenamefont {Jafri}, \citenamefont {Hou},
  \citenamefont {Wang}, \citenamefont {Ma},\ and\ \citenamefont
  {Wu}}]{liu2017transition}%
  \BibitemOpen
  \bibfield  {author} {\bibinfo {author} {\bibfnamefont {Z.}~\bibnamefont
  {Liu}}, \bibinfo {author} {\bibfnamefont {Y.}~\bibnamefont {Zhang}}, \bibinfo
  {author} {\bibfnamefont {G.}~\bibnamefont {Liu}}, \bibinfo {author}
  {\bibfnamefont {B.}~\bibnamefont {Ding}}, \bibinfo {author} {\bibfnamefont
  {E.}~\bibnamefont {Liu}}, \bibinfo {author} {\bibfnamefont {H.~M.}\
  \bibnamefont {Jafri}}, \bibinfo {author} {\bibfnamefont {Z.}~\bibnamefont
  {Hou}}, \bibinfo {author} {\bibfnamefont {W.}~\bibnamefont {Wang}}, \bibinfo
  {author} {\bibfnamefont {X.}~\bibnamefont {Ma}},\ and\ \bibinfo {author}
  {\bibfnamefont {G.}~\bibnamefont {Wu}},\ }\bibfield  {title} {\bibinfo
  {title} {{Transition from anomalous Hall effect to topological Hall effect in
  hexagonal non-collinear magnet $\rm{Mn_3Ga}$}},\ }\href@noop {} {\bibfield
  {journal} {\bibinfo  {journal} {Sci. Rep.}\ }\textbf {\bibinfo {volume}
  {7}},\ \bibinfo {pages} {515} (\bibinfo {year} {2017})}\BibitemShut {NoStop}%
\bibitem [{\citenamefont {S{\"u}rgers}\ \emph {et~al.}(2014)\citenamefont
  {S{\"u}rgers}, \citenamefont {Fischer}, \citenamefont {Winkel},\ and\
  \citenamefont {L{\"o}hneysen}}]{surgers2014large}%
  \BibitemOpen
  \bibfield  {author} {\bibinfo {author} {\bibfnamefont {C.}~\bibnamefont
  {S{\"u}rgers}}, \bibinfo {author} {\bibfnamefont {G.}~\bibnamefont
  {Fischer}}, \bibinfo {author} {\bibfnamefont {P.}~\bibnamefont {Winkel}},\
  and\ \bibinfo {author} {\bibfnamefont {H.~v.}\ \bibnamefont
  {L{\"o}hneysen}},\ }\bibfield  {title} {\bibinfo {title} {{Large topological
  Hall effect in the non-collinear phase of an antiferromagnet}},\ }\href@noop
  {} {\bibfield  {journal} {\bibinfo  {journal} {Nat. Commun.}\ }\textbf
  {\bibinfo {volume} {5}},\ \bibinfo {pages} {3400} (\bibinfo {year}
  {2014})}\BibitemShut {NoStop}%
\bibitem [{\citenamefont {Chen}\ \emph {et~al.}(2021)\citenamefont {Chen},
  \citenamefont {Tomita}, \citenamefont {Minami}, \citenamefont {Fu},
  \citenamefont {Koretsune}, \citenamefont {Kitatani}, \citenamefont
  {Muhammad}, \citenamefont {Nishio-Hamane}, \citenamefont {Ishii},
  \citenamefont {Ishii} \emph {et~al.}}]{chen2021anomalous}%
  \BibitemOpen
  \bibfield  {author} {\bibinfo {author} {\bibfnamefont {T.}~\bibnamefont
  {Chen}}, \bibinfo {author} {\bibfnamefont {T.}~\bibnamefont {Tomita}},
  \bibinfo {author} {\bibfnamefont {S.}~\bibnamefont {Minami}}, \bibinfo
  {author} {\bibfnamefont {M.}~\bibnamefont {Fu}}, \bibinfo {author}
  {\bibfnamefont {T.}~\bibnamefont {Koretsune}}, \bibinfo {author}
  {\bibfnamefont {M.}~\bibnamefont {Kitatani}}, \bibinfo {author}
  {\bibfnamefont {I.}~\bibnamefont {Muhammad}}, \bibinfo {author}
  {\bibfnamefont {D.}~\bibnamefont {Nishio-Hamane}}, \bibinfo {author}
  {\bibfnamefont {R.}~\bibnamefont {Ishii}}, \bibinfo {author} {\bibfnamefont
  {F.}~\bibnamefont {Ishii}}, \emph {et~al.},\ }\bibfield  {title} {\bibinfo
  {title} {{Anomalous transport due to Weyl fermions in the chiral
  antiferromagnets $\rm{Mn_3X}$, X= Sn, Ge}},\ }\href@noop {} {\bibfield
  {journal} {\bibinfo  {journal} {Nat. Commun.}\ }\textbf {\bibinfo {volume}
  {12}},\ \bibinfo {pages} {572} (\bibinfo {year} {2021})}\BibitemShut
  {NoStop}%
\bibitem [{\citenamefont {Deng}\ \emph {et~al.}(2020)\citenamefont {Deng},
  \citenamefont {Yu}, \citenamefont {Shi}, \citenamefont {Guo}, \citenamefont
  {Xu}, \citenamefont {Wang}, \citenamefont {Chen},\ and\ \citenamefont
  {Zhang}}]{deng2020quantum}%
  \BibitemOpen
  \bibfield  {author} {\bibinfo {author} {\bibfnamefont {Y.}~\bibnamefont
  {Deng}}, \bibinfo {author} {\bibfnamefont {Y.}~\bibnamefont {Yu}}, \bibinfo
  {author} {\bibfnamefont {M.~Z.}\ \bibnamefont {Shi}}, \bibinfo {author}
  {\bibfnamefont {Z.}~\bibnamefont {Guo}}, \bibinfo {author} {\bibfnamefont
  {Z.}~\bibnamefont {Xu}}, \bibinfo {author} {\bibfnamefont {J.}~\bibnamefont
  {Wang}}, \bibinfo {author} {\bibfnamefont {X.~H.}\ \bibnamefont {Chen}},\
  and\ \bibinfo {author} {\bibfnamefont {Y.}~\bibnamefont {Zhang}},\ }\bibfield
   {title} {\bibinfo {title} {{Quantum anomalous Hall effect in intrinsic
  magnetic topological insulator $\rm{MnBi_2Te_4}$}},\ }\href@noop {}
  {\bibfield  {journal} {\bibinfo  {journal} {Science}\ }\textbf {\bibinfo
  {volume} {367}},\ \bibinfo {pages} {895} (\bibinfo {year}
  {2020})}\BibitemShut {NoStop}%
\bibitem [{\citenamefont {Yang}\ \emph {et~al.}(2021)\citenamefont {Yang},
  \citenamefont {Xu}, \citenamefont {Zhu}, \citenamefont {Niu}, \citenamefont
  {Xu}, \citenamefont {Peng}, \citenamefont {Cheng}, \citenamefont {Jia},
  \citenamefont {Huang}, \citenamefont {Xu} \emph {et~al.}}]{yang2021odd}%
  \BibitemOpen
  \bibfield  {author} {\bibinfo {author} {\bibfnamefont {S.}~\bibnamefont
  {Yang}}, \bibinfo {author} {\bibfnamefont {X.}~\bibnamefont {Xu}}, \bibinfo
  {author} {\bibfnamefont {Y.}~\bibnamefont {Zhu}}, \bibinfo {author}
  {\bibfnamefont {R.}~\bibnamefont {Niu}}, \bibinfo {author} {\bibfnamefont
  {C.}~\bibnamefont {Xu}}, \bibinfo {author} {\bibfnamefont {Y.}~\bibnamefont
  {Peng}}, \bibinfo {author} {\bibfnamefont {X.}~\bibnamefont {Cheng}},
  \bibinfo {author} {\bibfnamefont {X.}~\bibnamefont {Jia}}, \bibinfo {author}
  {\bibfnamefont {Y.}~\bibnamefont {Huang}}, \bibinfo {author} {\bibfnamefont
  {X.}~\bibnamefont {Xu}}, \emph {et~al.},\ }\bibfield  {title} {\bibinfo
  {title} {{Odd-even layer-number effect and layer-dependent magnetic phase
  diagrams in $\rm{MnBi_2Te_4}$}},\ }\href@noop {} {\bibfield  {journal}
  {\bibinfo  {journal} {Phys. Rev. X}\ }\textbf {\bibinfo {volume} {11}},\
  \bibinfo {pages} {011003} (\bibinfo {year} {2021})}\BibitemShut {NoStop}%
\bibitem [{\citenamefont {Togawa}\ \emph {et~al.}(2012)\citenamefont {Togawa},
  \citenamefont {Koyama}, \citenamefont {Takayanagi}, \citenamefont {Mori},
  \citenamefont {Kousaka}, \citenamefont {Akimitsu}, \citenamefont {Nishihara},
  \citenamefont {Inoue}, \citenamefont {Ovchinnikov},\ and\ \citenamefont
  {Kishine}}]{togawa2012chiral}%
  \BibitemOpen
  \bibfield  {author} {\bibinfo {author} {\bibfnamefont {Y.}~\bibnamefont
  {Togawa}}, \bibinfo {author} {\bibfnamefont {T.}~\bibnamefont {Koyama}},
  \bibinfo {author} {\bibfnamefont {K.}~\bibnamefont {Takayanagi}}, \bibinfo
  {author} {\bibfnamefont {S.}~\bibnamefont {Mori}}, \bibinfo {author}
  {\bibfnamefont {Y.}~\bibnamefont {Kousaka}}, \bibinfo {author} {\bibfnamefont
  {J.}~\bibnamefont {Akimitsu}}, \bibinfo {author} {\bibfnamefont
  {S.}~\bibnamefont {Nishihara}}, \bibinfo {author} {\bibfnamefont
  {K.}~\bibnamefont {Inoue}}, \bibinfo {author} {\bibfnamefont
  {A.}~\bibnamefont {Ovchinnikov}},\ and\ \bibinfo {author} {\bibfnamefont
  {J.-i.}\ \bibnamefont {Kishine}},\ }\bibfield  {title} {\bibinfo {title}
  {Chiral magnetic soliton lattice on a chiral helimagnet},\ }\href@noop {}
  {\bibfield  {journal} {\bibinfo  {journal} {Phys. Rev. Lett.}\ }\textbf
  {\bibinfo {volume} {108}},\ \bibinfo {pages} {107202} (\bibinfo {year}
  {2012})}\BibitemShut {NoStop}%
\bibitem [{\citenamefont {Nair}\ \emph {et~al.}(2020)\citenamefont {Nair},
  \citenamefont {Maniv}, \citenamefont {John}, \citenamefont {Doyle},
  \citenamefont {Orenstein},\ and\ \citenamefont
  {Analytis}}]{nair2020electrical}%
  \BibitemOpen
  \bibfield  {author} {\bibinfo {author} {\bibfnamefont {N.~L.}\ \bibnamefont
  {Nair}}, \bibinfo {author} {\bibfnamefont {E.}~\bibnamefont {Maniv}},
  \bibinfo {author} {\bibfnamefont {C.}~\bibnamefont {John}}, \bibinfo {author}
  {\bibfnamefont {S.}~\bibnamefont {Doyle}}, \bibinfo {author} {\bibfnamefont
  {J.}~\bibnamefont {Orenstein}},\ and\ \bibinfo {author} {\bibfnamefont
  {J.~G.}\ \bibnamefont {Analytis}},\ }\bibfield  {title} {\bibinfo {title}
  {Electrical switching in a magnetically intercalated transition metal
  dichalcogenide},\ }\href@noop {} {\bibfield  {journal} {\bibinfo  {journal}
  {Nat. Mater.}\ }\textbf {\bibinfo {volume} {19}},\ \bibinfo {pages} {153}
  (\bibinfo {year} {2020})}\BibitemShut {NoStop}%
\bibitem [{\citenamefont {Maniv}\ \emph {et~al.}(2021)\citenamefont {Maniv},
  \citenamefont {Murphy}, \citenamefont {Haley}, \citenamefont {Doyle},
  \citenamefont {John}, \citenamefont {Maniv}, \citenamefont {Ramakrishna},
  \citenamefont {Tang}, \citenamefont {Ercius}, \citenamefont {Ramesh} \emph
  {et~al.}}]{maniv2021exchange}%
  \BibitemOpen
  \bibfield  {author} {\bibinfo {author} {\bibfnamefont {E.}~\bibnamefont
  {Maniv}}, \bibinfo {author} {\bibfnamefont {R.~A.}\ \bibnamefont {Murphy}},
  \bibinfo {author} {\bibfnamefont {S.~C.}\ \bibnamefont {Haley}}, \bibinfo
  {author} {\bibfnamefont {S.}~\bibnamefont {Doyle}}, \bibinfo {author}
  {\bibfnamefont {C.}~\bibnamefont {John}}, \bibinfo {author} {\bibfnamefont
  {A.}~\bibnamefont {Maniv}}, \bibinfo {author} {\bibfnamefont {S.~K.}\
  \bibnamefont {Ramakrishna}}, \bibinfo {author} {\bibfnamefont {Y.-L.}\
  \bibnamefont {Tang}}, \bibinfo {author} {\bibfnamefont {P.}~\bibnamefont
  {Ercius}}, \bibinfo {author} {\bibfnamefont {R.}~\bibnamefont {Ramesh}},
  \emph {et~al.},\ }\bibfield  {title} {\bibinfo {title} {Exchange bias due to
  coupling between coexisting antiferromagnetic and spin-glass orders},\
  }\href@noop {} {\bibfield  {journal} {\bibinfo  {journal} {Nat. Phys.}\
  }\textbf {\bibinfo {volume} {17}},\ \bibinfo {pages} {525} (\bibinfo {year}
  {2021})}\BibitemShut {NoStop}%
\bibitem [{\citenamefont {Ghimire}\ \emph {et~al.}(2018)\citenamefont
  {Ghimire}, \citenamefont {Botana}, \citenamefont {Jiang}, \citenamefont
  {Zhang}, \citenamefont {Chen},\ and\ \citenamefont
  {Mitchell}}]{ghimire2018large}%
  \BibitemOpen
  \bibfield  {author} {\bibinfo {author} {\bibfnamefont {N.~J.}\ \bibnamefont
  {Ghimire}}, \bibinfo {author} {\bibfnamefont {A.}~\bibnamefont {Botana}},
  \bibinfo {author} {\bibfnamefont {J.}~\bibnamefont {Jiang}}, \bibinfo
  {author} {\bibfnamefont {J.}~\bibnamefont {Zhang}}, \bibinfo {author}
  {\bibfnamefont {Y.-S.}\ \bibnamefont {Chen}},\ and\ \bibinfo {author}
  {\bibfnamefont {J.}~\bibnamefont {Mitchell}},\ }\bibfield  {title} {\bibinfo
  {title} {{Large anomalous Hall effect in the chiral-lattice antiferromagnet
  $\rm{CoNb_3S_6}$}},\ }\href@noop {} {\bibfield  {journal} {\bibinfo
  {journal} {Nat. Commun.}\ }\textbf {\bibinfo {volume} {9}},\ \bibinfo {pages}
  {3280} (\bibinfo {year} {2018})}\BibitemShut {NoStop}%
\bibitem [{\citenamefont {Tenasini}\ \emph {et~al.}(2020)\citenamefont
  {Tenasini}, \citenamefont {Martino}, \citenamefont {Ubrig}, \citenamefont
  {Ghimire}, \citenamefont {Berger}, \citenamefont {Zaharko}, \citenamefont
  {Wu}, \citenamefont {Mitchell}, \citenamefont {Martin}, \citenamefont
  {Forr{\'o}} \emph {et~al.}}]{tenasini2020giant}%
  \BibitemOpen
  \bibfield  {author} {\bibinfo {author} {\bibfnamefont {G.}~\bibnamefont
  {Tenasini}}, \bibinfo {author} {\bibfnamefont {E.}~\bibnamefont {Martino}},
  \bibinfo {author} {\bibfnamefont {N.}~\bibnamefont {Ubrig}}, \bibinfo
  {author} {\bibfnamefont {N.~J.}\ \bibnamefont {Ghimire}}, \bibinfo {author}
  {\bibfnamefont {H.}~\bibnamefont {Berger}}, \bibinfo {author} {\bibfnamefont
  {O.}~\bibnamefont {Zaharko}}, \bibinfo {author} {\bibfnamefont
  {F.}~\bibnamefont {Wu}}, \bibinfo {author} {\bibfnamefont {J.}~\bibnamefont
  {Mitchell}}, \bibinfo {author} {\bibfnamefont {I.}~\bibnamefont {Martin}},
  \bibinfo {author} {\bibfnamefont {L.}~\bibnamefont {Forr{\'o}}}, \emph
  {et~al.},\ }\bibfield  {title} {\bibinfo {title} {{Giant anomalous Hall
  effect in quasi-two-dimensional layered antiferromagnet $\rm{Co
  _{1/3}NbS_2}$}},\ }\href@noop {} {\bibfield  {journal} {\bibinfo  {journal}
  {Phys. Rev. Res.}\ }\textbf {\bibinfo {volume} {2}},\ \bibinfo {pages}
  {023051} (\bibinfo {year} {2020})}\BibitemShut {NoStop}%
\bibitem [{\citenamefont {Tanaka}\ \emph {et~al.}(2022)\citenamefont {Tanaka},
  \citenamefont {Okazaki}, \citenamefont {Kuroda}, \citenamefont {Noguchi},
  \citenamefont {Arai}, \citenamefont {Minami}, \citenamefont {Ideta},
  \citenamefont {Tanaka}, \citenamefont {Lu}, \citenamefont {Hashimoto} \emph
  {et~al.}}]{tanaka2022large}%
  \BibitemOpen
  \bibfield  {author} {\bibinfo {author} {\bibfnamefont {H.}~\bibnamefont
  {Tanaka}}, \bibinfo {author} {\bibfnamefont {S.}~\bibnamefont {Okazaki}},
  \bibinfo {author} {\bibfnamefont {K.}~\bibnamefont {Kuroda}}, \bibinfo
  {author} {\bibfnamefont {R.}~\bibnamefont {Noguchi}}, \bibinfo {author}
  {\bibfnamefont {Y.}~\bibnamefont {Arai}}, \bibinfo {author} {\bibfnamefont
  {S.}~\bibnamefont {Minami}}, \bibinfo {author} {\bibfnamefont
  {S.}~\bibnamefont {Ideta}}, \bibinfo {author} {\bibfnamefont
  {K.}~\bibnamefont {Tanaka}}, \bibinfo {author} {\bibfnamefont
  {D.}~\bibnamefont {Lu}}, \bibinfo {author} {\bibfnamefont {M.}~\bibnamefont
  {Hashimoto}}, \emph {et~al.},\ }\bibfield  {title} {\bibinfo {title} {{Large
  anomalous Hall effect induced by weak ferromagnetism in the
  noncentrosymmetric antiferromagnet $\rm{CoNb_3S_6}$}},\ }\href@noop {}
  {\bibfield  {journal} {\bibinfo  {journal} {Phys. Rev. B}\ }\textbf {\bibinfo
  {volume} {105}},\ \bibinfo {pages} {L121102} (\bibinfo {year}
  {2022})}\BibitemShut {NoStop}%
\bibitem [{\citenamefont {Yang}\ \emph {et~al.}(2022)\citenamefont {Yang},
  \citenamefont {LaBollita}, \citenamefont {Cheng}, \citenamefont {Bhandari},
  \citenamefont {Cochran}, \citenamefont {Yin}, \citenamefont {Hossain},
  \citenamefont {Belopolski}, \citenamefont {Zhang}, \citenamefont {Jiang}
  \emph {et~al.}}]{yang2022visualizing}%
  \BibitemOpen
  \bibfield  {author} {\bibinfo {author} {\bibfnamefont {X.~P.}\ \bibnamefont
  {Yang}}, \bibinfo {author} {\bibfnamefont {H.}~\bibnamefont {LaBollita}},
  \bibinfo {author} {\bibfnamefont {Z.-J.}\ \bibnamefont {Cheng}}, \bibinfo
  {author} {\bibfnamefont {H.}~\bibnamefont {Bhandari}}, \bibinfo {author}
  {\bibfnamefont {T.~A.}\ \bibnamefont {Cochran}}, \bibinfo {author}
  {\bibfnamefont {J.-X.}\ \bibnamefont {Yin}}, \bibinfo {author} {\bibfnamefont
  {M.~S.}\ \bibnamefont {Hossain}}, \bibinfo {author} {\bibfnamefont
  {I.}~\bibnamefont {Belopolski}}, \bibinfo {author} {\bibfnamefont
  {Q.}~\bibnamefont {Zhang}}, \bibinfo {author} {\bibfnamefont
  {Y.}~\bibnamefont {Jiang}}, \emph {et~al.},\ }\bibfield  {title} {\bibinfo
  {title} {Visualizing the out-of-plane electronic dispersions in an
  intercalated transition metal dichalcogenide},\ }\href@noop {} {\bibfield
  {journal} {\bibinfo  {journal} {Phys. Rev. B}\ }\textbf {\bibinfo {volume}
  {105}},\ \bibinfo {pages} {L121107} (\bibinfo {year} {2022})}\BibitemShut
  {NoStop}%
\bibitem [{\citenamefont {Pop{\v{c}}evi{\'c}}\ \emph
  {et~al.}(2022)\citenamefont {Pop{\v{c}}evi{\'c}}, \citenamefont {Utsumi},
  \citenamefont {Bia{\l}o}, \citenamefont {Tabis}, \citenamefont {Gala},
  \citenamefont {Rosmus}, \citenamefont {Kolodziej}, \citenamefont
  {Tomaszewska}, \citenamefont {Garb}, \citenamefont {Berger} \emph
  {et~al.}}]{popvcevic2022role}%
  \BibitemOpen
  \bibfield  {author} {\bibinfo {author} {\bibfnamefont {P.}~\bibnamefont
  {Pop{\v{c}}evi{\'c}}}, \bibinfo {author} {\bibfnamefont {Y.}~\bibnamefont
  {Utsumi}}, \bibinfo {author} {\bibfnamefont {I.}~\bibnamefont {Bia{\l}o}},
  \bibinfo {author} {\bibfnamefont {W.}~\bibnamefont {Tabis}}, \bibinfo
  {author} {\bibfnamefont {M.~A.}\ \bibnamefont {Gala}}, \bibinfo {author}
  {\bibfnamefont {M.}~\bibnamefont {Rosmus}}, \bibinfo {author} {\bibfnamefont
  {J.~J.}\ \bibnamefont {Kolodziej}}, \bibinfo {author} {\bibfnamefont
  {N.}~\bibnamefont {Tomaszewska}}, \bibinfo {author} {\bibfnamefont
  {M.}~\bibnamefont {Garb}}, \bibinfo {author} {\bibfnamefont {H.}~\bibnamefont
  {Berger}}, \emph {et~al.},\ }\bibfield  {title} {\bibinfo {title} {{Role of
  intercalated cobalt in the electronic structure of $\rm{Co _{1/3}NbS_2}$}},\
  }\href@noop {} {\bibfield  {journal} {\bibinfo  {journal} {Phys. Rev. B}\
  }\textbf {\bibinfo {volume} {105}},\ \bibinfo {pages} {155114} (\bibinfo
  {year} {2022})}\BibitemShut {NoStop}%
\bibitem [{\citenamefont {Mangelsen}\ \emph {et~al.}(2021)\citenamefont
  {Mangelsen}, \citenamefont {Zimmer}, \citenamefont {N{\"a}ther},
  \citenamefont {Mankovsky}, \citenamefont {Polesya}, \citenamefont {Ebert},\
  and\ \citenamefont {Bensch}}]{mangelsen2021interplay}%
  \BibitemOpen
  \bibfield  {author} {\bibinfo {author} {\bibfnamefont {S.}~\bibnamefont
  {Mangelsen}}, \bibinfo {author} {\bibfnamefont {P.}~\bibnamefont {Zimmer}},
  \bibinfo {author} {\bibfnamefont {C.}~\bibnamefont {N{\"a}ther}}, \bibinfo
  {author} {\bibfnamefont {S.}~\bibnamefont {Mankovsky}}, \bibinfo {author}
  {\bibfnamefont {S.}~\bibnamefont {Polesya}}, \bibinfo {author} {\bibfnamefont
  {H.}~\bibnamefont {Ebert}},\ and\ \bibinfo {author} {\bibfnamefont
  {W.}~\bibnamefont {Bensch}},\ }\bibfield  {title} {\bibinfo {title}
  {{Interplay of sample composition and anomalous Hall effect in $\rm{Co
  _{x}NbS_2}$}},\ }\href@noop {} {\bibfield  {journal} {\bibinfo  {journal}
  {Phys. Rev. B}\ }\textbf {\bibinfo {volume} {103}},\ \bibinfo {pages}
  {184408} (\bibinfo {year} {2021})}\BibitemShut {NoStop}%
\bibitem [{\citenamefont {Parkin}\ \emph {et~al.}(1983)\citenamefont {Parkin},
  \citenamefont {Marseglia},\ and\ \citenamefont {Brown}}]{parkin1983magnetic}%
  \BibitemOpen
  \bibfield  {author} {\bibinfo {author} {\bibfnamefont {S.}~\bibnamefont
  {Parkin}}, \bibinfo {author} {\bibfnamefont {E.}~\bibnamefont {Marseglia}},\
  and\ \bibinfo {author} {\bibfnamefont {P.}~\bibnamefont {Brown}},\ }\bibfield
   {title} {\bibinfo {title} {{Magnetic structure of $\rm{Co _{1/3}NbS_2}$ and
  $\rm{Co _{1/3}TaS_2}$}},\ }\href@noop {} {\bibfield  {journal} {\bibinfo
  {journal} {J. Phys. C}\ }\textbf {\bibinfo {volume} {16}},\ \bibinfo {pages}
  {2765} (\bibinfo {year} {1983})}\BibitemShut {NoStop}%
\bibitem [{\citenamefont {Zhang}\ \emph {et~al.}(2023)\citenamefont {Zhang},
  \citenamefont {Deng}, \citenamefont {Sheng}, \citenamefont {Liu},
  \citenamefont {Kumar}, \citenamefont {Shimada}, \citenamefont {Jiang},
  \citenamefont {Liu}, \citenamefont {Shen}, \citenamefont {Li} \emph
  {et~al.}}]{zhang2023chiral}%
  \BibitemOpen
  \bibfield  {author} {\bibinfo {author} {\bibfnamefont {A.}~\bibnamefont
  {Zhang}}, \bibinfo {author} {\bibfnamefont {K.}~\bibnamefont {Deng}},
  \bibinfo {author} {\bibfnamefont {J.}~\bibnamefont {Sheng}}, \bibinfo
  {author} {\bibfnamefont {P.}~\bibnamefont {Liu}}, \bibinfo {author}
  {\bibfnamefont {S.}~\bibnamefont {Kumar}}, \bibinfo {author} {\bibfnamefont
  {K.}~\bibnamefont {Shimada}}, \bibinfo {author} {\bibfnamefont
  {Z.}~\bibnamefont {Jiang}}, \bibinfo {author} {\bibfnamefont
  {Z.}~\bibnamefont {Liu}}, \bibinfo {author} {\bibfnamefont {D.}~\bibnamefont
  {Shen}}, \bibinfo {author} {\bibfnamefont {J.}~\bibnamefont {Li}}, \emph
  {et~al.},\ }\bibfield  {title} {\bibinfo {title} {{Chiral Dirac fermion in a
  collinear antiferromagnet}},\ }\href@noop {} {\bibfield  {journal} {\bibinfo
  {journal} {arXiv preprint arXiv:2301.12201}\ } (\bibinfo {year}
  {2023})}\BibitemShut {NoStop}%
\bibitem [{\citenamefont {{\v{S}}mejkal}\ \emph {et~al.}(2020)\citenamefont
  {{\v{S}}mejkal}, \citenamefont {Gonz{\'a}lez-Hern{\'a}ndez}, \citenamefont
  {Jungwirth},\ and\ \citenamefont {Sinova}}]{vsmejkal2020crystal}%
  \BibitemOpen
  \bibfield  {author} {\bibinfo {author} {\bibfnamefont {L.}~\bibnamefont
  {{\v{S}}mejkal}}, \bibinfo {author} {\bibfnamefont {R.}~\bibnamefont
  {Gonz{\'a}lez-Hern{\'a}ndez}}, \bibinfo {author} {\bibfnamefont
  {T.}~\bibnamefont {Jungwirth}},\ and\ \bibinfo {author} {\bibfnamefont
  {J.}~\bibnamefont {Sinova}},\ }\bibfield  {title} {\bibinfo {title} {{Crystal
  time-reversal symmetry breaking and spontaneous Hall effect in collinear
  antiferromagnets}},\ }\href@noop {} {\bibfield  {journal} {\bibinfo
  {journal} {Sci. Adv.}\ }\textbf {\bibinfo {volume} {6}},\ \bibinfo {pages}
  {eaaz8809} (\bibinfo {year} {2020})}\BibitemShut {NoStop}%
\bibitem [{\citenamefont {Lu}\ \emph {et~al.}(2022)\citenamefont {Lu},
  \citenamefont {Murzabekova}, \citenamefont {Shim}, \citenamefont {Park},
  \citenamefont {Kim}, \citenamefont {Kish}, \citenamefont {Wu}, \citenamefont
  {DeBeer-Schmitt}, \citenamefont {Aczel}, \citenamefont {Schleife} \emph
  {et~al.}}]{lu2022understanding}%
  \BibitemOpen
  \bibfield  {author} {\bibinfo {author} {\bibfnamefont {K.}~\bibnamefont
  {Lu}}, \bibinfo {author} {\bibfnamefont {A.}~\bibnamefont {Murzabekova}},
  \bibinfo {author} {\bibfnamefont {S.}~\bibnamefont {Shim}}, \bibinfo {author}
  {\bibfnamefont {J.}~\bibnamefont {Park}}, \bibinfo {author} {\bibfnamefont
  {S.}~\bibnamefont {Kim}}, \bibinfo {author} {\bibfnamefont {L.}~\bibnamefont
  {Kish}}, \bibinfo {author} {\bibfnamefont {Y.}~\bibnamefont {Wu}}, \bibinfo
  {author} {\bibfnamefont {L.}~\bibnamefont {DeBeer-Schmitt}}, \bibinfo
  {author} {\bibfnamefont {A.}~\bibnamefont {Aczel}}, \bibinfo {author}
  {\bibfnamefont {A.}~\bibnamefont {Schleife}}, \emph {et~al.},\ }\bibfield
  {title} {\bibinfo {title} {{Understanding the Anomalous Hall effect in
  $\rm{Co_{1/3}NbS_2}$ from crystal and magnetic structures}},\ }\href@noop {}
  {\bibfield  {journal} {\bibinfo  {journal} {arXiv preprint arXiv:2212.14762}\
  } (\bibinfo {year} {2022})}\BibitemShut {NoStop}%
\bibitem [{\citenamefont {Park}\ \emph {et~al.}(2022)\citenamefont {Park},
  \citenamefont {Heinonen},\ and\ \citenamefont {Martin}}]{park2022first}%
  \BibitemOpen
  \bibfield  {author} {\bibinfo {author} {\bibfnamefont {H.}~\bibnamefont
  {Park}}, \bibinfo {author} {\bibfnamefont {O.}~\bibnamefont {Heinonen}},\
  and\ \bibinfo {author} {\bibfnamefont {I.}~\bibnamefont {Martin}},\
  }\bibfield  {title} {\bibinfo {title} {{First-principles study of magnetic
  states and the anomalous Hall conductivity of $\rm{MNb_3S_6}$ (M = Co, Fe,
  Mn, and Ni)}},\ }\href@noop {} {\bibfield  {journal} {\bibinfo  {journal}
  {Phys. Rev. Mater.}\ }\textbf {\bibinfo {volume} {6}},\ \bibinfo {pages}
  {024201} (\bibinfo {year} {2022})}\BibitemShut {NoStop}%
\bibitem [{\citenamefont {Heinonen}\ \emph {et~al.}(2022)\citenamefont
  {Heinonen}, \citenamefont {Heinonen},\ and\ \citenamefont
  {Park}}]{heinonen2022magnetic}%
  \BibitemOpen
  \bibfield  {author} {\bibinfo {author} {\bibfnamefont {O.}~\bibnamefont
  {Heinonen}}, \bibinfo {author} {\bibfnamefont {R.}~\bibnamefont {Heinonen}},\
  and\ \bibinfo {author} {\bibfnamefont {H.}~\bibnamefont {Park}},\ }\bibfield
  {title} {\bibinfo {title} {{Magnetic ground states of a model for
  $\rm{MNb_3S_6}$ (M = Co, Fe, and Ni)}},\ }\href@noop {} {\bibfield  {journal}
  {\bibinfo  {journal} {Phys. Rev. Mater}\ }\textbf {\bibinfo {volume} {6}},\
  \bibinfo {pages} {024405} (\bibinfo {year} {2022})}\BibitemShut {NoStop}%
\bibitem [{\citenamefont {Takagi}\ \emph {et~al.}(2023)\citenamefont {Takagi},
  \citenamefont {Takagi}, \citenamefont {Minami}, \citenamefont {Nomoto},
  \citenamefont {Ohishi}, \citenamefont {Suzuki}, \citenamefont {Yanagi},
  \citenamefont {Hirayama}, \citenamefont {Khanh}, \citenamefont {Karube} \emph
  {et~al.}}]{takagi2023spontaneous}%
  \BibitemOpen
  \bibfield  {author} {\bibinfo {author} {\bibfnamefont {H.}~\bibnamefont
  {Takagi}}, \bibinfo {author} {\bibfnamefont {R.}~\bibnamefont {Takagi}},
  \bibinfo {author} {\bibfnamefont {S.}~\bibnamefont {Minami}}, \bibinfo
  {author} {\bibfnamefont {T.}~\bibnamefont {Nomoto}}, \bibinfo {author}
  {\bibfnamefont {K.}~\bibnamefont {Ohishi}}, \bibinfo {author} {\bibfnamefont
  {M.-T.}\ \bibnamefont {Suzuki}}, \bibinfo {author} {\bibfnamefont
  {Y.}~\bibnamefont {Yanagi}}, \bibinfo {author} {\bibfnamefont
  {M.}~\bibnamefont {Hirayama}}, \bibinfo {author} {\bibfnamefont
  {N.}~\bibnamefont {Khanh}}, \bibinfo {author} {\bibfnamefont
  {K.}~\bibnamefont {Karube}}, \emph {et~al.},\ }\bibfield  {title} {\bibinfo
  {title} {{Spontaneous topological Hall effect induced by non-coplanar
  antiferromagnetic order in intercalated van der Waals materials}},\
  }\href@noop {} {\bibfield  {journal} {\bibinfo  {journal} {Nat. Phys.}\ ,\
  \bibinfo {pages} {1}} (\bibinfo {year} {2023})}\BibitemShut {NoStop}%
\bibitem [{\citenamefont {Friend}\ \emph {et~al.}(1977)\citenamefont {Friend},
  \citenamefont {Beal},\ and\ \citenamefont {Yoffe}}]{friend1977electrical}%
  \BibitemOpen
  \bibfield  {author} {\bibinfo {author} {\bibfnamefont {R.}~\bibnamefont
  {Friend}}, \bibinfo {author} {\bibfnamefont {A.}~\bibnamefont {Beal}},\ and\
  \bibinfo {author} {\bibfnamefont {A.}~\bibnamefont {Yoffe}},\ }\bibfield
  {title} {\bibinfo {title} {Electrical and magnetic properties of some first
  row transition metal intercalates of niobium disulphide},\ }\href@noop {}
  {\bibfield  {journal} {\bibinfo  {journal} {Philos. Mag.}\ }\textbf {\bibinfo
  {volume} {35}},\ \bibinfo {pages} {1269} (\bibinfo {year}
  {1977})}\BibitemShut {NoStop}%
\bibitem [{\citenamefont {Suzuki}\ \emph {et~al.}(2017)\citenamefont {Suzuki},
  \citenamefont {Koretsune}, \citenamefont {Ochi},\ and\ \citenamefont
  {Arita}}]{suzuki2017cluster}%
  \BibitemOpen
  \bibfield  {author} {\bibinfo {author} {\bibfnamefont {M.-T.}\ \bibnamefont
  {Suzuki}}, \bibinfo {author} {\bibfnamefont {T.}~\bibnamefont {Koretsune}},
  \bibinfo {author} {\bibfnamefont {M.}~\bibnamefont {Ochi}},\ and\ \bibinfo
  {author} {\bibfnamefont {R.}~\bibnamefont {Arita}},\ }\bibfield  {title}
  {\bibinfo {title} {{Cluster multipole theory for anomalous Hall effect in
  antiferromagnets}},\ }\href@noop {} {\bibfield  {journal} {\bibinfo
  {journal} {Phys. Rev. B}\ }\textbf {\bibinfo {volume} {95}},\ \bibinfo
  {pages} {094406} (\bibinfo {year} {2017})}\BibitemShut {NoStop}%
\bibitem [{\citenamefont {Feng}\ \emph {et~al.}(2015)\citenamefont {Feng},
  \citenamefont {Guo}, \citenamefont {Zhou}, \citenamefont {Yao},\ and\
  \citenamefont {Niu}}]{feng2015large}%
  \BibitemOpen
  \bibfield  {author} {\bibinfo {author} {\bibfnamefont {W.}~\bibnamefont
  {Feng}}, \bibinfo {author} {\bibfnamefont {G.-Y.}\ \bibnamefont {Guo}},
  \bibinfo {author} {\bibfnamefont {J.}~\bibnamefont {Zhou}}, \bibinfo {author}
  {\bibfnamefont {Y.}~\bibnamefont {Yao}},\ and\ \bibinfo {author}
  {\bibfnamefont {Q.}~\bibnamefont {Niu}},\ }\bibfield  {title} {\bibinfo
  {title} {{Large magneto-optical Kerr effect in noncollinear antiferromagnets
  $\rm{Mn_3X}$ (X= Rh, Ir, Pt)}},\ }\href@noop {} {\bibfield  {journal}
  {\bibinfo  {journal} {Phys. Rev. B}\ }\textbf {\bibinfo {volume} {92}},\
  \bibinfo {pages} {144426} (\bibinfo {year} {2015})}\BibitemShut {NoStop}%
\bibitem [{\citenamefont {Higo}\ \emph {et~al.}(2018)\citenamefont {Higo},
  \citenamefont {Man}, \citenamefont {Gopman}, \citenamefont {Wu},
  \citenamefont {Koretsune}, \citenamefont {van’t Erve}, \citenamefont
  {Kabanov}, \citenamefont {Rees}, \citenamefont {Li}, \citenamefont {Suzuki}
  \emph {et~al.}}]{higo2018large}%
  \BibitemOpen
  \bibfield  {author} {\bibinfo {author} {\bibfnamefont {T.}~\bibnamefont
  {Higo}}, \bibinfo {author} {\bibfnamefont {H.}~\bibnamefont {Man}}, \bibinfo
  {author} {\bibfnamefont {D.~B.}\ \bibnamefont {Gopman}}, \bibinfo {author}
  {\bibfnamefont {L.}~\bibnamefont {Wu}}, \bibinfo {author} {\bibfnamefont
  {T.}~\bibnamefont {Koretsune}}, \bibinfo {author} {\bibfnamefont {O.~M.}\
  \bibnamefont {van’t Erve}}, \bibinfo {author} {\bibfnamefont {Y.~P.}\
  \bibnamefont {Kabanov}}, \bibinfo {author} {\bibfnamefont {D.}~\bibnamefont
  {Rees}}, \bibinfo {author} {\bibfnamefont {Y.}~\bibnamefont {Li}}, \bibinfo
  {author} {\bibfnamefont {M.-T.}\ \bibnamefont {Suzuki}}, \emph {et~al.},\
  }\bibfield  {title} {\bibinfo {title} {{Large magneto-optical Kerr effect and
  imaging of magnetic octupole domains in an antiferromagnetic metal}},\
  }\href@noop {} {\bibfield  {journal} {\bibinfo  {journal} {Nat. Photonics}\
  }\textbf {\bibinfo {volume} {12}},\ \bibinfo {pages} {73} (\bibinfo {year}
  {2018})}\BibitemShut {NoStop}%
\bibitem [{\citenamefont {Onoda}\ \emph {et~al.}(2008)\citenamefont {Onoda},
  \citenamefont {Sugimoto},\ and\ \citenamefont {Nagaosa}}]{onoda2008quantum}%
  \BibitemOpen
  \bibfield  {author} {\bibinfo {author} {\bibfnamefont {S.}~\bibnamefont
  {Onoda}}, \bibinfo {author} {\bibfnamefont {N.}~\bibnamefont {Sugimoto}},\
  and\ \bibinfo {author} {\bibfnamefont {N.}~\bibnamefont {Nagaosa}},\
  }\bibfield  {title} {\bibinfo {title} {{Quantum transport theory of anomalous
  electric, thermoelectric, and thermal Hall effects in ferromagnets}},\
  }\href@noop {} {\bibfield  {journal} {\bibinfo  {journal} {Phys. Rev. B}\
  }\textbf {\bibinfo {volume} {77}},\ \bibinfo {pages} {165103} (\bibinfo
  {year} {2008})}\BibitemShut {NoStop}%
\bibitem [{\citenamefont {Kim}\ \emph {et~al.}(2018)\citenamefont {Kim},
  \citenamefont {Seo}, \citenamefont {Lee}, \citenamefont {Ko}, \citenamefont
  {Kim}, \citenamefont {Jang}, \citenamefont {Ok}, \citenamefont {Lee},
  \citenamefont {Jo}, \citenamefont {Kang} \emph {et~al.}}]{kim2018large}%
  \BibitemOpen
  \bibfield  {author} {\bibinfo {author} {\bibfnamefont {K.}~\bibnamefont
  {Kim}}, \bibinfo {author} {\bibfnamefont {J.}~\bibnamefont {Seo}}, \bibinfo
  {author} {\bibfnamefont {E.}~\bibnamefont {Lee}}, \bibinfo {author}
  {\bibfnamefont {K.-T.}\ \bibnamefont {Ko}}, \bibinfo {author} {\bibfnamefont
  {B.}~\bibnamefont {Kim}}, \bibinfo {author} {\bibfnamefont {B.~G.}\
  \bibnamefont {Jang}}, \bibinfo {author} {\bibfnamefont {J.~M.}\ \bibnamefont
  {Ok}}, \bibinfo {author} {\bibfnamefont {J.}~\bibnamefont {Lee}}, \bibinfo
  {author} {\bibfnamefont {Y.~J.}\ \bibnamefont {Jo}}, \bibinfo {author}
  {\bibfnamefont {W.}~\bibnamefont {Kang}}, \emph {et~al.},\ }\bibfield
  {title} {\bibinfo {title} {{Large anomalous Hall current induced by
  topological nodal lines in a ferromagnetic van der Waals semimetal}},\
  }\href@noop {} {\bibfield  {journal} {\bibinfo  {journal} {Nat. Mater.}\
  }\textbf {\bibinfo {volume} {17}},\ \bibinfo {pages} {794} (\bibinfo {year}
  {2018})}\BibitemShut {NoStop}%
\bibitem [{\citenamefont {Xiao}\ \emph {et~al.}(2006)\citenamefont {Xiao},
  \citenamefont {Yao}, \citenamefont {Fang},\ and\ \citenamefont
  {Niu}}]{xiao2006berry}%
  \BibitemOpen
  \bibfield  {author} {\bibinfo {author} {\bibfnamefont {D.}~\bibnamefont
  {Xiao}}, \bibinfo {author} {\bibfnamefont {Y.}~\bibnamefont {Yao}}, \bibinfo
  {author} {\bibfnamefont {Z.}~\bibnamefont {Fang}},\ and\ \bibinfo {author}
  {\bibfnamefont {Q.}~\bibnamefont {Niu}},\ }\bibfield  {title} {\bibinfo
  {title} {Berry-phase effect in anomalous thermoelectric transport},\
  }\href@noop {} {\bibfield  {journal} {\bibinfo  {journal} {Phys. Rev. Lett.}\
  }\textbf {\bibinfo {volume} {97}},\ \bibinfo {pages} {026603} (\bibinfo
  {year} {2006})}\BibitemShut {NoStop}%
\bibitem [{\citenamefont {Wuttke}\ \emph {et~al.}(2019)\citenamefont {Wuttke},
  \citenamefont {Caglieris}, \citenamefont {Sykora}, \citenamefont
  {Scaravaggi}, \citenamefont {Wolter}, \citenamefont {Manna}, \citenamefont
  {S{\"u}ss}, \citenamefont {Shekhar}, \citenamefont {Felser}, \citenamefont
  {B{\"u}chner} \emph {et~al.}}]{wuttke2019berry}%
  \BibitemOpen
  \bibfield  {author} {\bibinfo {author} {\bibfnamefont {C.}~\bibnamefont
  {Wuttke}}, \bibinfo {author} {\bibfnamefont {F.}~\bibnamefont {Caglieris}},
  \bibinfo {author} {\bibfnamefont {S.}~\bibnamefont {Sykora}}, \bibinfo
  {author} {\bibfnamefont {F.}~\bibnamefont {Scaravaggi}}, \bibinfo {author}
  {\bibfnamefont {A.~U.}\ \bibnamefont {Wolter}}, \bibinfo {author}
  {\bibfnamefont {K.}~\bibnamefont {Manna}}, \bibinfo {author} {\bibfnamefont
  {V.}~\bibnamefont {S{\"u}ss}}, \bibinfo {author} {\bibfnamefont
  {C.}~\bibnamefont {Shekhar}}, \bibinfo {author} {\bibfnamefont
  {C.}~\bibnamefont {Felser}}, \bibinfo {author} {\bibfnamefont
  {B.}~\bibnamefont {B{\"u}chner}}, \emph {et~al.},\ }\bibfield  {title}
  {\bibinfo {title} {{Berry curvature unravelled by the anomalous Nernst effect
  in $\rm{Mn_3Ge}$}},\ }\href@noop {} {\bibfield  {journal} {\bibinfo
  {journal} {Phys. Rev. B}\ }\textbf {\bibinfo {volume} {100}},\ \bibinfo
  {pages} {085111} (\bibinfo {year} {2019})}\BibitemShut {NoStop}%
\bibitem [{\citenamefont {Bari{\v{s}}i{\'c}}\ \emph {et~al.}(2011)\citenamefont
  {Bari{\v{s}}i{\'c}}, \citenamefont {Smiljani{\'c}}, \citenamefont
  {Pop{\v{c}}evi{\'c}}, \citenamefont {Bilu{\v{s}}i{\'c}}, \citenamefont
  {Tuti{\v{s}}}, \citenamefont {Smontara}, \citenamefont {Berger},
  \citenamefont {Ja{\'c}imovi{\'c}}, \citenamefont {Yuli},\ and\ \citenamefont
  {Forr{\'o}}}]{barivsic2011high}%
  \BibitemOpen
  \bibfield  {author} {\bibinfo {author} {\bibfnamefont {N.}~\bibnamefont
  {Bari{\v{s}}i{\'c}}}, \bibinfo {author} {\bibfnamefont {I.}~\bibnamefont
  {Smiljani{\'c}}}, \bibinfo {author} {\bibfnamefont {P.}~\bibnamefont
  {Pop{\v{c}}evi{\'c}}}, \bibinfo {author} {\bibfnamefont {A.}~\bibnamefont
  {Bilu{\v{s}}i{\'c}}}, \bibinfo {author} {\bibfnamefont {E.}~\bibnamefont
  {Tuti{\v{s}}}}, \bibinfo {author} {\bibfnamefont {A.}~\bibnamefont
  {Smontara}}, \bibinfo {author} {\bibfnamefont {H.}~\bibnamefont {Berger}},
  \bibinfo {author} {\bibfnamefont {J.}~\bibnamefont {Ja{\'c}imovi{\'c}}},
  \bibinfo {author} {\bibfnamefont {O.}~\bibnamefont {Yuli}},\ and\ \bibinfo
  {author} {\bibfnamefont {L.}~\bibnamefont {Forr{\'o}}},\ }\bibfield  {title}
  {\bibinfo {title} {{High-pressure study of transport properties in
  $\rm{Co_{0.33}NbS_ 2}$}},\ }\href@noop {} {\bibfield  {journal} {\bibinfo
  {journal} {Phys. Rev. B}\ }\textbf {\bibinfo {volume} {84}},\ \bibinfo
  {pages} {075157} (\bibinfo {year} {2011})}\BibitemShut {NoStop}%
\bibitem [{\citenamefont {Inoue}\ \emph {et~al.}(1986)\citenamefont {Inoue},
  \citenamefont {Muneta}, \citenamefont {Negishi},\ and\ \citenamefont
  {Sasaki}}]{inoue1986specific}%
  \BibitemOpen
  \bibfield  {author} {\bibinfo {author} {\bibfnamefont {M.}~\bibnamefont
  {Inoue}}, \bibinfo {author} {\bibfnamefont {Y.}~\bibnamefont {Muneta}},
  \bibinfo {author} {\bibfnamefont {H.}~\bibnamefont {Negishi}},\ and\ \bibinfo
  {author} {\bibfnamefont {M.}~\bibnamefont {Sasaki}},\ }\bibfield  {title}
  {\bibinfo {title} {{Specific heat measurements of intercalation compounds
  $\rm{M_xTiS_2}$ (M = 3$d$ transition metals) using ac calorimetry
  technique}},\ }\href@noop {} {\bibfield  {journal} {\bibinfo  {journal} {J.
  Low Temp. Phys.}\ }\textbf {\bibinfo {volume} {63}},\ \bibinfo {pages} {235}
  (\bibinfo {year} {1986})}\BibitemShut {NoStop}%
\bibitem [{\citenamefont {Pop{\v{c}}evi{\'c}}\ \emph
  {et~al.}(2020)\citenamefont {Pop{\v{c}}evi{\'c}}, \citenamefont
  {Batisti{\'c}}, \citenamefont {Smontara}, \citenamefont {Velebit},
  \citenamefont {Ja{\'c}imovi{\'c}}, \citenamefont {Martino}, \citenamefont
  {{\v{Z}}ivkovi{\'c}}, \citenamefont {Tsyrulin}, \citenamefont {Piatek},
  \citenamefont {Berger} \emph {et~al.}}]{popvcevic2020electronic}%
  \BibitemOpen
  \bibfield  {author} {\bibinfo {author} {\bibfnamefont {P.}~\bibnamefont
  {Pop{\v{c}}evi{\'c}}}, \bibinfo {author} {\bibfnamefont {I.}~\bibnamefont
  {Batisti{\'c}}}, \bibinfo {author} {\bibfnamefont {A.}~\bibnamefont
  {Smontara}}, \bibinfo {author} {\bibfnamefont {K.}~\bibnamefont {Velebit}},
  \bibinfo {author} {\bibfnamefont {J.}~\bibnamefont {Ja{\'c}imovi{\'c}}},
  \bibinfo {author} {\bibfnamefont {E.}~\bibnamefont {Martino}}, \bibinfo
  {author} {\bibfnamefont {I.}~\bibnamefont {{\v{Z}}ivkovi{\'c}}}, \bibinfo
  {author} {\bibfnamefont {N.}~\bibnamefont {Tsyrulin}}, \bibinfo {author}
  {\bibfnamefont {J.}~\bibnamefont {Piatek}}, \bibinfo {author} {\bibfnamefont
  {H.}~\bibnamefont {Berger}}, \emph {et~al.},\ }\bibfield  {title} {\bibinfo
  {title} {{Electronic transport and magnetism in the alternating stack of
  metallic and highly frustrated magnetic layers in $\rm{Co_{1/3}NbS_2}$}},\
  }\href@noop {} {\bibfield  {journal} {\bibinfo  {journal} {arXiv preprint
  arXiv:2003.08127}\ } (\bibinfo {year} {2020})}\BibitemShut {NoStop}%
\bibitem [{\citenamefont {Xu}\ \emph {et~al.}(2019)\citenamefont {Xu},
  \citenamefont {Phelan},\ and\ \citenamefont {Chien}}]{xu2019large}%
  \BibitemOpen
  \bibfield  {author} {\bibinfo {author} {\bibfnamefont {J.}~\bibnamefont
  {Xu}}, \bibinfo {author} {\bibfnamefont {W.~A.}\ \bibnamefont {Phelan}},\
  and\ \bibinfo {author} {\bibfnamefont {C.-L.}\ \bibnamefont {Chien}},\
  }\bibfield  {title} {\bibinfo {title} {{Large anomalous Nernst effect in a
  van der Waals ferromagnet $\rm{Fe_3GeTe_2}$}},\ }\href@noop {} {\bibfield
  {journal} {\bibinfo  {journal} {Nano Lett.}\ }\textbf {\bibinfo {volume}
  {19}},\ \bibinfo {pages} {8250} (\bibinfo {year} {2019})}\BibitemShut
  {NoStop}%
\bibitem [{\citenamefont {Hasegawa}\ \emph {et~al.}(2015)\citenamefont
  {Hasegawa}, \citenamefont {Mizuguchi}, \citenamefont {Sakuraba},
  \citenamefont {Kamada}, \citenamefont {Kojima}, \citenamefont {Kubota},
  \citenamefont {Mizukami}, \citenamefont {Miyazaki},\ and\ \citenamefont
  {Takanashi}}]{hasegawa2015material}%
  \BibitemOpen
  \bibfield  {author} {\bibinfo {author} {\bibfnamefont {K.}~\bibnamefont
  {Hasegawa}}, \bibinfo {author} {\bibfnamefont {M.}~\bibnamefont {Mizuguchi}},
  \bibinfo {author} {\bibfnamefont {Y.}~\bibnamefont {Sakuraba}}, \bibinfo
  {author} {\bibfnamefont {T.}~\bibnamefont {Kamada}}, \bibinfo {author}
  {\bibfnamefont {T.}~\bibnamefont {Kojima}}, \bibinfo {author} {\bibfnamefont
  {T.}~\bibnamefont {Kubota}}, \bibinfo {author} {\bibfnamefont
  {S.}~\bibnamefont {Mizukami}}, \bibinfo {author} {\bibfnamefont
  {T.}~\bibnamefont {Miyazaki}},\ and\ \bibinfo {author} {\bibfnamefont
  {K.}~\bibnamefont {Takanashi}},\ }\bibfield  {title} {\bibinfo {title}
  {{Material dependence of anomalous Nernst effect in perpendicularly
  magnetized ordered-alloy thin films}},\ }\href@noop {} {\bibfield  {journal}
  {\bibinfo  {journal} {Appl. Phys. Lett.}\ }\textbf {\bibinfo {volume}
  {106}},\ \bibinfo {pages} {252405} (\bibinfo {year} {2015})}\BibitemShut
  {NoStop}%
\bibitem [{\citenamefont {Chuang}\ \emph {et~al.}(2017)\citenamefont {Chuang},
  \citenamefont {Su}, \citenamefont {Wu},\ and\ \citenamefont
  {Huang}}]{chuang2017enhancement}%
  \BibitemOpen
  \bibfield  {author} {\bibinfo {author} {\bibfnamefont {T.-C.}\ \bibnamefont
  {Chuang}}, \bibinfo {author} {\bibfnamefont {P.}~\bibnamefont {Su}}, \bibinfo
  {author} {\bibfnamefont {P.}~\bibnamefont {Wu}},\ and\ \bibinfo {author}
  {\bibfnamefont {S.~Y.}\ \bibnamefont {Huang}},\ }\bibfield  {title} {\bibinfo
  {title} {{Enhancement of the anomalous Nernst effect in ferromagnetic thin
  films}},\ }\href@noop {} {\bibfield  {journal} {\bibinfo  {journal} {Phys.
  Rev. B}\ }\textbf {\bibinfo {volume} {96}},\ \bibinfo {pages} {174406}
  (\bibinfo {year} {2017})}\BibitemShut {NoStop}%
\bibitem [{\citenamefont {Sakai}\ \emph {et~al.}(2018)\citenamefont {Sakai},
  \citenamefont {Mizuta}, \citenamefont {Nugroho}, \citenamefont {Sihombing},
  \citenamefont {Koretsune}, \citenamefont {Suzuki}, \citenamefont {Takemori},
  \citenamefont {Ishii}, \citenamefont {Nishio-Hamane}, \citenamefont {Arita}
  \emph {et~al.}}]{sakai2018giant}%
  \BibitemOpen
  \bibfield  {author} {\bibinfo {author} {\bibfnamefont {A.}~\bibnamefont
  {Sakai}}, \bibinfo {author} {\bibfnamefont {Y.~P.}\ \bibnamefont {Mizuta}},
  \bibinfo {author} {\bibfnamefont {A.~A.}\ \bibnamefont {Nugroho}}, \bibinfo
  {author} {\bibfnamefont {R.}~\bibnamefont {Sihombing}}, \bibinfo {author}
  {\bibfnamefont {T.}~\bibnamefont {Koretsune}}, \bibinfo {author}
  {\bibfnamefont {M.-T.}\ \bibnamefont {Suzuki}}, \bibinfo {author}
  {\bibfnamefont {N.}~\bibnamefont {Takemori}}, \bibinfo {author}
  {\bibfnamefont {R.}~\bibnamefont {Ishii}}, \bibinfo {author} {\bibfnamefont
  {D.}~\bibnamefont {Nishio-Hamane}}, \bibinfo {author} {\bibfnamefont
  {R.}~\bibnamefont {Arita}}, \emph {et~al.},\ }\bibfield  {title} {\bibinfo
  {title} {{Giant anomalous Nernst effect and quantum-critical scaling in a
  ferromagnetic semimetal}},\ }\href@noop {} {\bibfield  {journal} {\bibinfo
  {journal} {Nat. Phys.}\ }\textbf {\bibinfo {volume} {14}},\ \bibinfo {pages}
  {1119} (\bibinfo {year} {2018})}\BibitemShut {NoStop}%
\bibitem [{\citenamefont {Yang}\ \emph {et~al.}(2020)\citenamefont {Yang},
  \citenamefont {You}, \citenamefont {Wang}, \citenamefont {Huang},
  \citenamefont {Xi}, \citenamefont {Xu}, \citenamefont {Cao}, \citenamefont
  {Tian}, \citenamefont {Xu}, \citenamefont {Dai} \emph
  {et~al.}}]{yang2020giant}%
  \BibitemOpen
  \bibfield  {author} {\bibinfo {author} {\bibfnamefont {H.}~\bibnamefont
  {Yang}}, \bibinfo {author} {\bibfnamefont {W.}~\bibnamefont {You}}, \bibinfo
  {author} {\bibfnamefont {J.}~\bibnamefont {Wang}}, \bibinfo {author}
  {\bibfnamefont {J.}~\bibnamefont {Huang}}, \bibinfo {author} {\bibfnamefont
  {C.}~\bibnamefont {Xi}}, \bibinfo {author} {\bibfnamefont {X.}~\bibnamefont
  {Xu}}, \bibinfo {author} {\bibfnamefont {C.}~\bibnamefont {Cao}}, \bibinfo
  {author} {\bibfnamefont {M.}~\bibnamefont {Tian}}, \bibinfo {author}
  {\bibfnamefont {Z.-A.}\ \bibnamefont {Xu}}, \bibinfo {author} {\bibfnamefont
  {J.}~\bibnamefont {Dai}}, \emph {et~al.},\ }\bibfield  {title} {\bibinfo
  {title} {{Giant anomalous Nernst effect in the magnetic Weyl semimetal
  $\rm{Co_3Sn_2S_2}$}},\ }\href@noop {} {\bibfield  {journal} {\bibinfo
  {journal} {Phys. Rev. Mater.}\ }\textbf {\bibinfo {volume} {4}},\ \bibinfo
  {pages} {024202} (\bibinfo {year} {2020})}\BibitemShut {NoStop}%
\bibitem [{\citenamefont {Pan}\ \emph {et~al.}(2022)\citenamefont {Pan},
  \citenamefont {Le}, \citenamefont {He}, \citenamefont {Watzman},
  \citenamefont {Yao}, \citenamefont {Gooth}, \citenamefont {Heremans},
  \citenamefont {Sun},\ and\ \citenamefont {Felser}}]{pan2022giant}%
  \BibitemOpen
  \bibfield  {author} {\bibinfo {author} {\bibfnamefont {Y.}~\bibnamefont
  {Pan}}, \bibinfo {author} {\bibfnamefont {C.}~\bibnamefont {Le}}, \bibinfo
  {author} {\bibfnamefont {B.}~\bibnamefont {He}}, \bibinfo {author}
  {\bibfnamefont {S.~J.}\ \bibnamefont {Watzman}}, \bibinfo {author}
  {\bibfnamefont {M.}~\bibnamefont {Yao}}, \bibinfo {author} {\bibfnamefont
  {J.}~\bibnamefont {Gooth}}, \bibinfo {author} {\bibfnamefont {J.~P.}\
  \bibnamefont {Heremans}}, \bibinfo {author} {\bibfnamefont {Y.}~\bibnamefont
  {Sun}},\ and\ \bibinfo {author} {\bibfnamefont {C.}~\bibnamefont {Felser}},\
  }\bibfield  {title} {\bibinfo {title} {{Giant anomalous Nernst signal in the
  antiferromagnet $\rm{YbMnBi_2}$}},\ }\href@noop {} {\bibfield  {journal}
  {\bibinfo  {journal} {Nat. Commun.}\ }\textbf {\bibinfo {volume} {21}},\
  \bibinfo {pages} {203} (\bibinfo {year} {2022})}\BibitemShut {NoStop}%
\bibitem [{\citenamefont {Pu}\ \emph {et~al.}(2008)\citenamefont {Pu},
  \citenamefont {Chiba}, \citenamefont {Matsukura}, \citenamefont {Ohno},\ and\
  \citenamefont {Shi}}]{pu2008mott}%
  \BibitemOpen
  \bibfield  {author} {\bibinfo {author} {\bibfnamefont {Y.}~\bibnamefont
  {Pu}}, \bibinfo {author} {\bibfnamefont {D.}~\bibnamefont {Chiba}}, \bibinfo
  {author} {\bibfnamefont {F.}~\bibnamefont {Matsukura}}, \bibinfo {author}
  {\bibfnamefont {H.}~\bibnamefont {Ohno}},\ and\ \bibinfo {author}
  {\bibfnamefont {J.}~\bibnamefont {Shi}},\ }\bibfield  {title} {\bibinfo
  {title} {{Mott relation for anomalous Hall and Nernst effects in
  $\rm{Ga_{1-x}Mn_xAs}$ ferromagnetic semiconductors}},\ }\href@noop {}
  {\bibfield  {journal} {\bibinfo  {journal} {Phys. Rev. Lett.}\ }\textbf
  {\bibinfo {volume} {101}},\ \bibinfo {pages} {117208} (\bibinfo {year}
  {2008})}\BibitemShut {NoStop}%
\bibitem [{\citenamefont {Sun}\ \emph {et~al.}(2019)\citenamefont {Sun},
  \citenamefont {Yi}, \citenamefont {Song}, \citenamefont {Clark},
  \citenamefont {Huang}, \citenamefont {Shan}, \citenamefont {Wu},
  \citenamefont {Huang}, \citenamefont {Gao}, \citenamefont {Chen} \emph
  {et~al.}}]{sun2019giant}%
  \BibitemOpen
  \bibfield  {author} {\bibinfo {author} {\bibfnamefont {Z.}~\bibnamefont
  {Sun}}, \bibinfo {author} {\bibfnamefont {Y.}~\bibnamefont {Yi}}, \bibinfo
  {author} {\bibfnamefont {T.}~\bibnamefont {Song}}, \bibinfo {author}
  {\bibfnamefont {G.}~\bibnamefont {Clark}}, \bibinfo {author} {\bibfnamefont
  {B.}~\bibnamefont {Huang}}, \bibinfo {author} {\bibfnamefont
  {Y.}~\bibnamefont {Shan}}, \bibinfo {author} {\bibfnamefont {S.}~\bibnamefont
  {Wu}}, \bibinfo {author} {\bibfnamefont {D.}~\bibnamefont {Huang}}, \bibinfo
  {author} {\bibfnamefont {C.}~\bibnamefont {Gao}}, \bibinfo {author}
  {\bibfnamefont {Z.}~\bibnamefont {Chen}}, \emph {et~al.},\ }\bibfield
  {title} {\bibinfo {title} {{Giant nonreciprocal second-harmonic generation
  from antiferromagnetic bilayer $\rm{CrI_3}$}},\ }\href@noop {} {\bibfield
  {journal} {\bibinfo  {journal} {Nature}\ }\textbf {\bibinfo {volume} {572}},\
  \bibinfo {pages} {497} (\bibinfo {year} {2019})}\BibitemShut {NoStop}%
\bibitem [{\citenamefont {Ni}\ \emph {et~al.}(2021)\citenamefont {Ni},
  \citenamefont {Zhang}, \citenamefont {Hopper}, \citenamefont {Haglund},
  \citenamefont {Huang}, \citenamefont {Jariwala}, \citenamefont {Bassett},
  \citenamefont {Mandrus}, \citenamefont {Mele}, \citenamefont {Kane} \emph
  {et~al.}}]{ni2021direct}%
  \BibitemOpen
  \bibfield  {author} {\bibinfo {author} {\bibfnamefont {Z.}~\bibnamefont
  {Ni}}, \bibinfo {author} {\bibfnamefont {H.}~\bibnamefont {Zhang}}, \bibinfo
  {author} {\bibfnamefont {D.~A.}\ \bibnamefont {Hopper}}, \bibinfo {author}
  {\bibfnamefont {A.~V.}\ \bibnamefont {Haglund}}, \bibinfo {author}
  {\bibfnamefont {N.}~\bibnamefont {Huang}}, \bibinfo {author} {\bibfnamefont
  {D.}~\bibnamefont {Jariwala}}, \bibinfo {author} {\bibfnamefont {L.~C.}\
  \bibnamefont {Bassett}}, \bibinfo {author} {\bibfnamefont {D.~G.}\
  \bibnamefont {Mandrus}}, \bibinfo {author} {\bibfnamefont {E.~J.}\
  \bibnamefont {Mele}}, \bibinfo {author} {\bibfnamefont {C.~L.}\ \bibnamefont
  {Kane}}, \emph {et~al.},\ }\bibfield  {title} {\bibinfo {title} {{Direct
  imaging of antiferromagnetic domains and anomalous layer-dependent mirror
  symmetry breaking in atomically thin $\rm{MnPS_3}$}},\ }\href@noop {}
  {\bibfield  {journal} {\bibinfo  {journal} {Physical review letters}\
  }\textbf {\bibinfo {volume} {127}},\ \bibinfo {pages} {187201} (\bibinfo
  {year} {2021})}\BibitemShut {NoStop}%
\bibitem [{\citenamefont {Schumacher}(1991)}]{schumacher1991modification}%
  \BibitemOpen
  \bibfield  {author} {\bibinfo {author} {\bibfnamefont {F.}~\bibnamefont
  {Schumacher}},\ }\bibfield  {title} {\bibinfo {title} {{On the modification
  of the Kondorsky function}},\ }\href@noop {} {\bibfield  {journal} {\bibinfo
  {journal} {J. Appl. Phys.}\ }\textbf {\bibinfo {volume} {70}},\ \bibinfo
  {pages} {3184} (\bibinfo {year} {1991})}\BibitemShut {NoStop}%
\bibitem [{\citenamefont {Stoner}\ and\ \citenamefont
  {Wohlfarth}(1948)}]{stoner1948mechanism}%
  \BibitemOpen
  \bibfield  {author} {\bibinfo {author} {\bibfnamefont {E.~C.}\ \bibnamefont
  {Stoner}}\ and\ \bibinfo {author} {\bibfnamefont {E.}~\bibnamefont
  {Wohlfarth}},\ }\bibfield  {title} {\bibinfo {title} {A mechanism of magnetic
  hysteresis in heterogeneous alloys},\ }\href@noop {} {\bibfield  {journal}
  {\bibinfo  {journal} {Philos. Trans. R. Soc. Lond. A}\ }\textbf {\bibinfo
  {volume} {240}},\ \bibinfo {pages} {599} (\bibinfo {year}
  {1948})}\BibitemShut {NoStop}%
\bibitem [{\citenamefont {Li}\ \emph {et~al.}(2019)\citenamefont {Li},
  \citenamefont {Collignon}, \citenamefont {Xu}, \citenamefont {Zuo},
  \citenamefont {Cavanna}, \citenamefont {Gennser}, \citenamefont {Mailly},
  \citenamefont {Fauqu{\'e}}, \citenamefont {Balents}, \citenamefont {Zhu}
  \emph {et~al.}}]{li2019chiral}%
  \BibitemOpen
  \bibfield  {author} {\bibinfo {author} {\bibfnamefont {X.}~\bibnamefont
  {Li}}, \bibinfo {author} {\bibfnamefont {C.}~\bibnamefont {Collignon}},
  \bibinfo {author} {\bibfnamefont {L.}~\bibnamefont {Xu}}, \bibinfo {author}
  {\bibfnamefont {H.}~\bibnamefont {Zuo}}, \bibinfo {author} {\bibfnamefont
  {A.}~\bibnamefont {Cavanna}}, \bibinfo {author} {\bibfnamefont
  {U.}~\bibnamefont {Gennser}}, \bibinfo {author} {\bibfnamefont
  {D.}~\bibnamefont {Mailly}}, \bibinfo {author} {\bibfnamefont
  {B.}~\bibnamefont {Fauqu{\'e}}}, \bibinfo {author} {\bibfnamefont
  {L.}~\bibnamefont {Balents}}, \bibinfo {author} {\bibfnamefont
  {Z.}~\bibnamefont {Zhu}}, \emph {et~al.},\ }\bibfield  {title} {\bibinfo
  {title} {{Chiral domain walls of $\rm{Mn_3Sn}$ and their memory}},\
  }\href@noop {} {\bibfield  {journal} {\bibinfo  {journal} {Nat. Commun.}\
  }\textbf {\bibinfo {volume} {10}},\ \bibinfo {pages} {3021} (\bibinfo {year}
  {2019})}\BibitemShut {NoStop}%
\end{thebibliography}%
	
	\setcounter{figure}{0}
	\setcounter{table}{0}
	\newpage

	\nocite{*}
	
\end{document}


\date{}
\onecolumn{
\maketitle 
\vspace{-5mm}
\begin{center}
\begin{minipage}{1\textwidth}
\begin{center}

\textit{
\textsuperscript{1} State Key Laboratory for Artificial Microstructure $\rm{\&}$ Mesoscopic Physics and Frontiers Science Center for Nano-Optoelectronics, School of Physics, Peking University, Beijing 100871, China
\\\textsuperscript{2} Academy for Advanced Interdisciplinary Studies, Peking University, Beijing 100871, China
\\\textsuperscript{3} Department of Physics and Beijing Key Laboratory of Optoelectronic Functional Materials and Micro-Nano Devices, Renmin University of China, Beijing 100872, China
\\\textsuperscript{4} Shenyang National Laboratory for Materials Science, Institute of Metal Research, Chinese Academy of Sciences, Shenyang 110016, China
\\\textsuperscript{5} Key Laboratory of Quantum State Construction and Manipulation (Ministry of Education), Renmin University of China, Beijing, 100872, China
\\\textsuperscript{6} Laboratory for Neutron Scattering, Remin University of China, Beijing, 100872, China
\\\textsuperscript{7} Collaboration International Center of Quantum Matter, Beijing 100871, China
\\\textsuperscript{8} Liaoning Academy of Materials, Shenyang, 110167, China
\\{$\dagger$} Emails: tlxia@ruc.edu.cn, jbyang@pku.edu.cn, ye\_yu@pku.edu.cn\\
\vspace{5mm}
}

\end{center}
\end{minipage}
\end{center}

\clearpage
\newpage 

\setlength\parindent{12pt}
\noindent 

\begin{figure*}[ht!]
\centering
\includegraphics[width=0.65\textwidth]{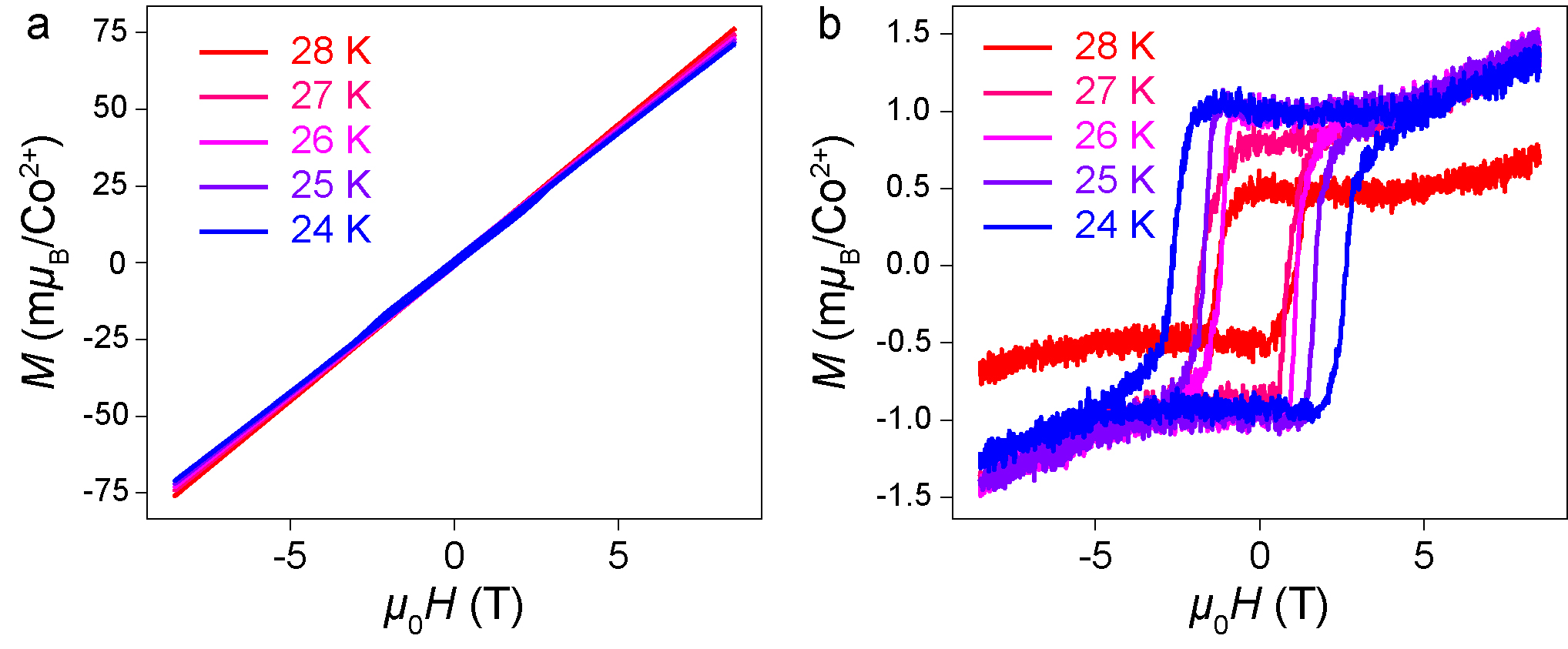}
\caption{\textbf{Magnetization measurements of single crystal $\rm{Co_{1/3}NbS_2}$ bulk.} 
\textbf{(a),} $M-H$ curves measured at different temperatures with the magnetic field applied along the $c$ axis. The magnetization in $\rm{Co_{1/3}NbS_2}$ linearly increases with the external field. \textbf{(b),} The hysteresis loops of the $M-H$ curves after subtracting the linear background, extracted from (a).
}

\label{MH}
\end{figure*}

\begin{figure*}[ht!]
	\centering
	\includegraphics[width=0.95\textwidth]{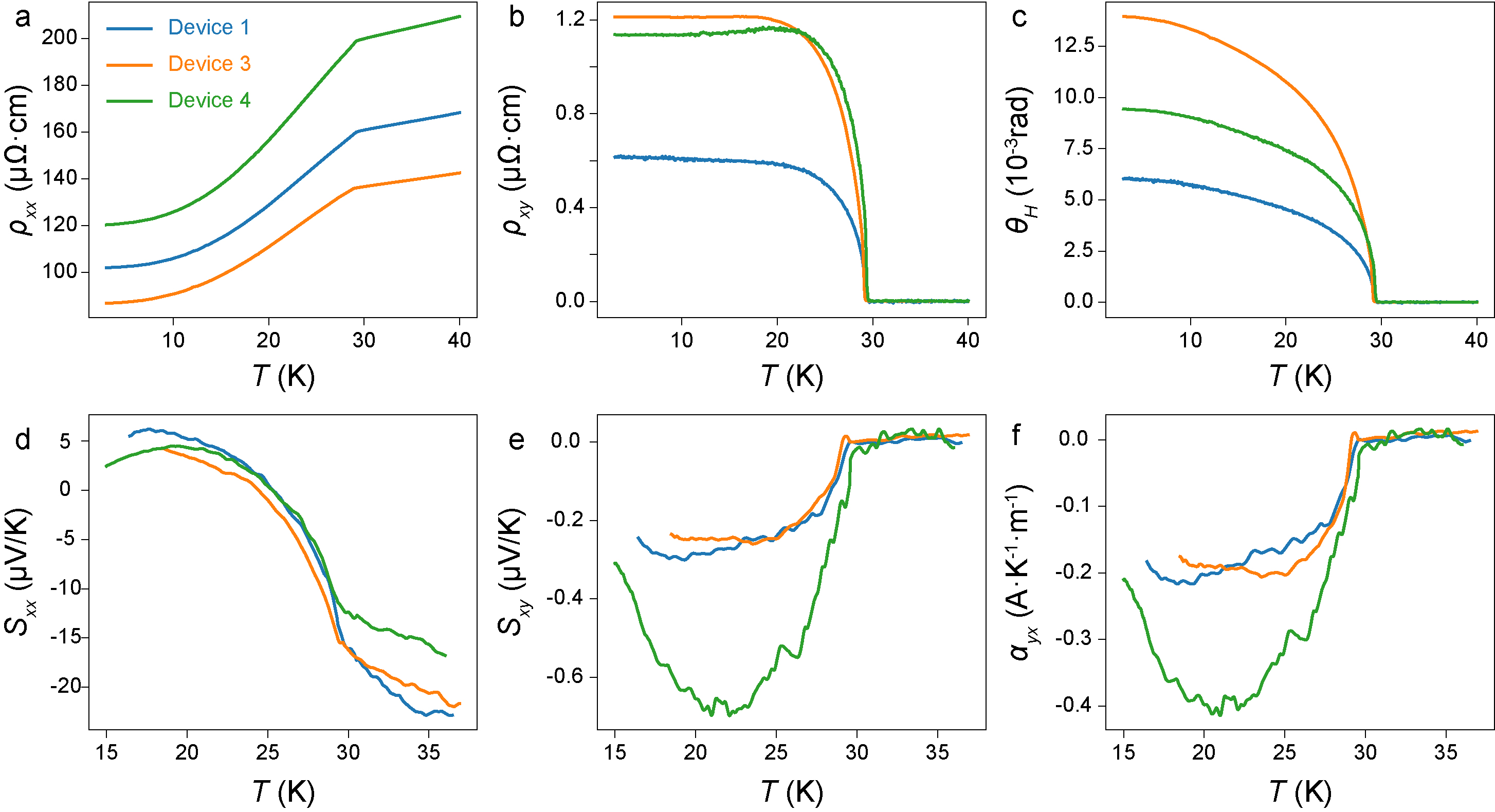}
	\caption{\textbf{Transport and thermoelectric measurements of different devices.} 
		\textbf{(a, b, c),} $\rho_{xx}$ (a), $\rho_{xy}$ (b) and the calculated Hall angle $\theta_H$ on the dependence of temperature. The blue, orange, and blue curves label the measurement of three different devices, namely device 1 (125 nm), device 3 (110 nm), and device 4 (40 nm), respectively. \textbf{(d, e, f),} $S_{xx}$ (d), $S_{xy}$ (e) and the calculated thermoelectric conductivity $\alpha_{yx}$ (f) on the dependence of temperature.}
	\label{Devices}
\end{figure*}

\begin{table}[ht!]
\centering
\caption{Thickness ($t$), residual resistivity ratio (RRR), maximum anomalous Hall conductivity ($\sigma_{yx}^A$), maximum Hall angle ($\theta_H$), maximum anomalous Nernst coefficient ($S_{yx}$), Nernst angle at the lowest temperature ($\theta_N$) and maximum transverse thermoelectric conductivity ($\alpha_{xy}$) of the three measured devices.}
\renewcommand\arraystretch{2}
\begin{tabular}{lccccccc}
	\hline
	& $t$ (nm) & RRR    & $\sigma_{yx}^A\ \rm{(\upOmega^{-1}\cdot cm^{-1})}$ & $\theta_H\ \rm{(10^{-3})}$ & $S_{yx}\ \rm{(\upmu V/K)}$  & $\theta_N$ & $\alpha_{xy}\ \rm{(A\cdot K^{-1}\cdot m^{-1})}$ \\ \hline
	Device 1 & 125    & 3.0257 & 59.45            & 6.07     & 0.301 & 0.0443   & 0.217     \\
	Device 3 & 40     & 2.9250 & 160.58           & 13.96    & 0.260 & 0.0558   & 0.206     \\
	Device 4 & 110    & 3.2027 & 78.47            & 9.44     & 0.698 & 0.1254   & 0.415     \\ \hline
\end{tabular}
\end{table}

\clearpage
\newpage 

\begin{figure*}[ht!]
	\centering
	\includegraphics[width=0.7\textwidth]{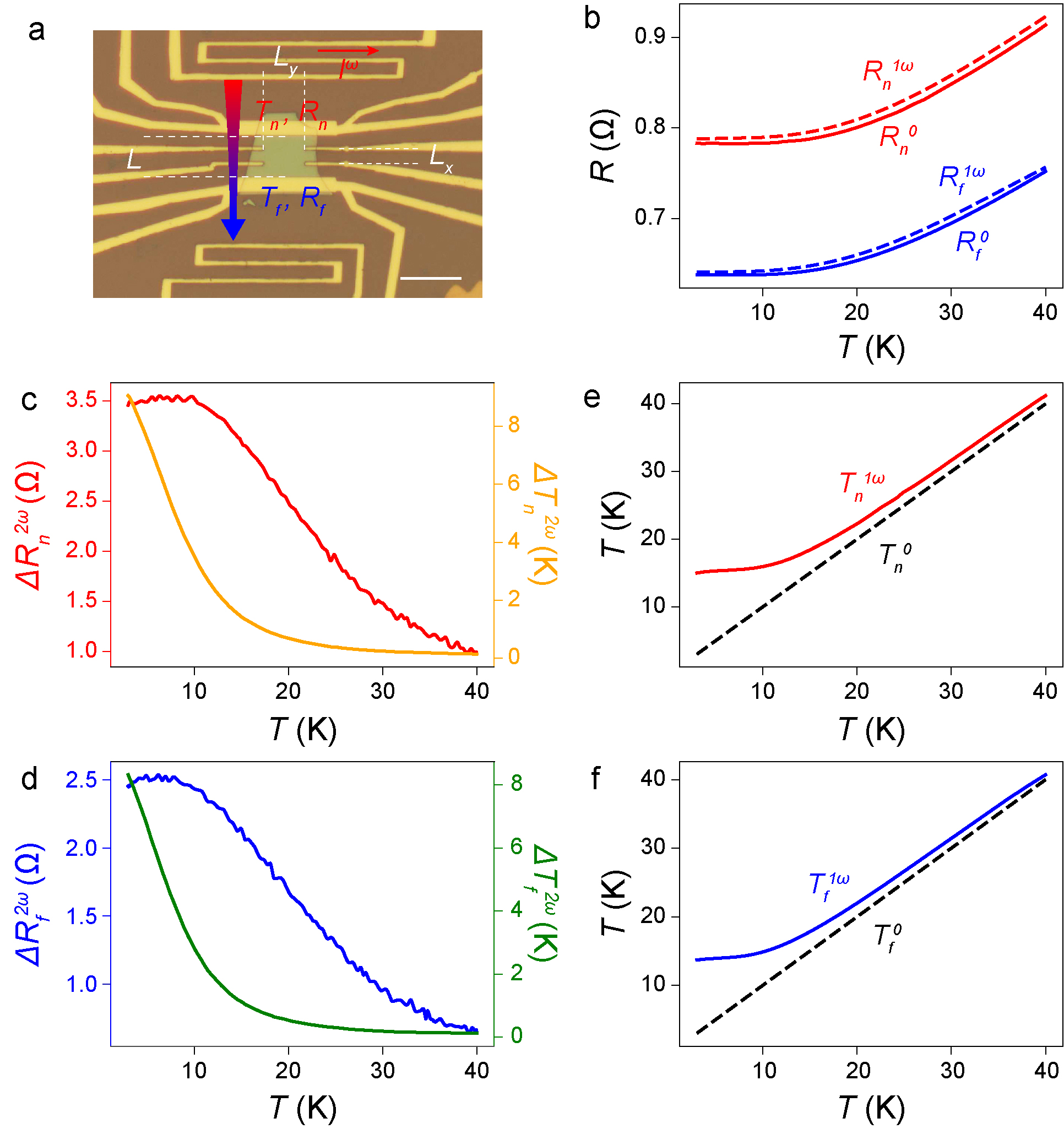}
	\caption{\textbf{Calibration of the temperature gradient in thermoelectric devices.} 
		\textbf{(a),} Optical image of device 4 ($\sim$110 nm). The scale bar is 10 $\rm{\upmu m}$. \textbf{(b),} Four-probe resistance of the near (red) and far (blue) sensors \textit{versus} temperature. The solid and dashed lines show the resistances with the heater on and off, respectively. \textbf{(c, d),} The $2\omega$ resistance and the calculated $2\omega$ temperature perturbation of the near (c) and far sensor (d). \textbf{(e, f),} The calculated actual temperature $T^{1\omega}$ at the near (e) and the far (f) sensor. The dashed lines show the read-out temperature $T^0$.
	}
	
	\label{Calibration_1}
\end{figure*}

\begin{figure*}[ht!]
	\centering
	\includegraphics[width=0.95\textwidth]{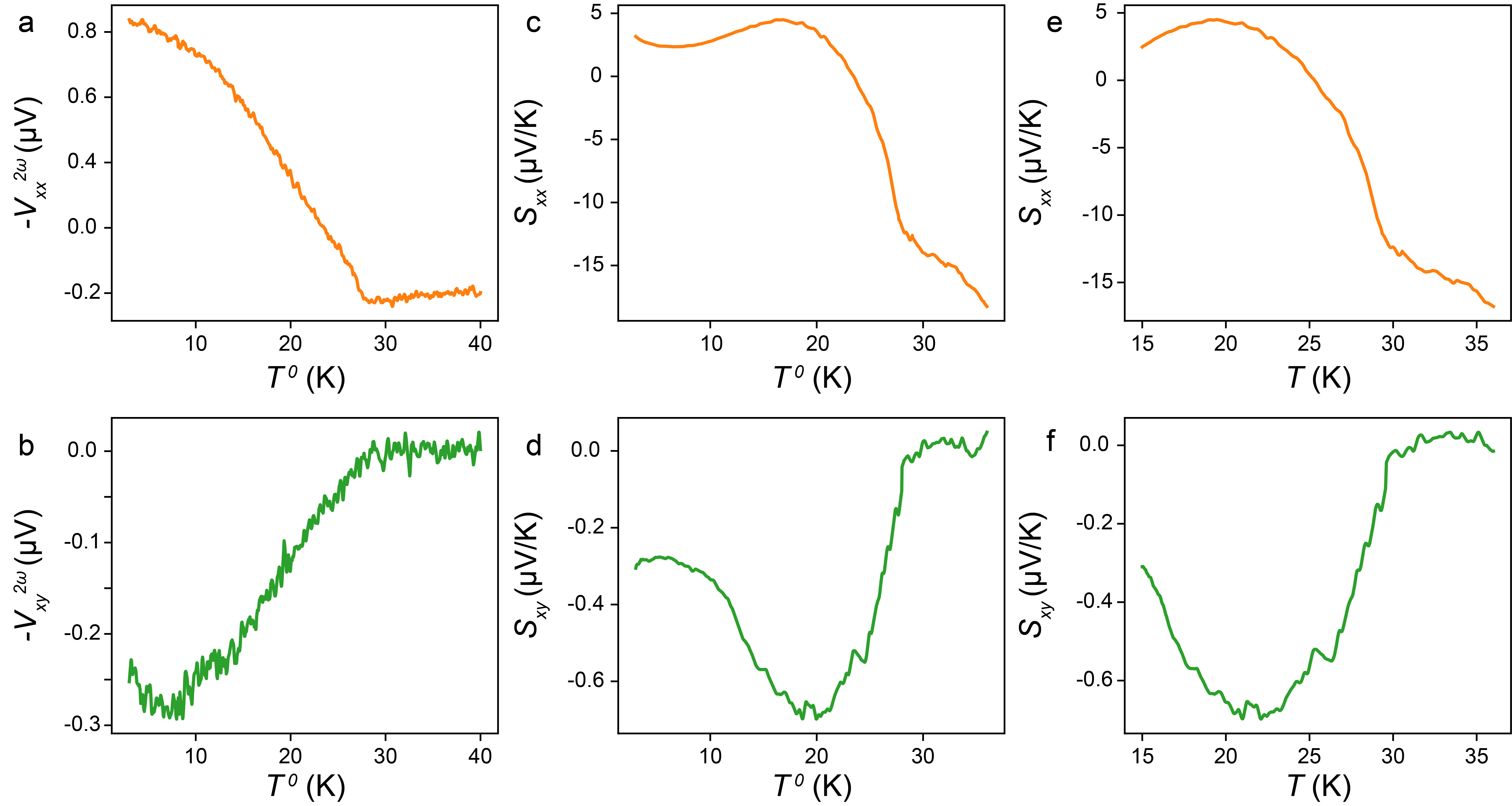}
	\caption{\textbf{Thermoelectric data processing of device 4.} 
		\textbf{(a, b),} Original longitudinal (a) and transverse (b) $2\omega$ thermoelectric signal \textit{versus} temperature. \textbf{(c,d),} Calculated Seebeck (c) and Nernst (d) coefficients \textit{versus} temperature by dividing the $2\omega$ temperature perturbation $(T_n^{2\omega}-T_f^{2\omega})$ and the scale factor ($L/L_x$ or $L/L_y$) extracted from Fig. \ref{Calibration_1}. \textbf{(e, f),} Calculated Seebeck (e) and Nernst (f) coefficients \textit{versus} calibrated actual temperature.
	}
	
	\label{Calibration_2}
\end{figure*}

Here, we demonstrate the calibration of the temperature gradient, the actual temperature, and the calculation of the thermoelectric coefficients in our devices. We take device 4 (110 nm) as an example. An a.c. current of 0.2 mA and 3.777 Hz was applied to the upper heater, generating a constant temperature rise and a $2\omega$ temperature perturbation. The temperature at the two sensors, which we denote as the near sensor and the far sensor according to their distances to the upper heater, can be expressed as:

\begin{equation}
\begin{split}
T_n = T_n^{1\omega}+\Delta T_n^{2\omega} = T_n^{0}+\Delta T_n^{1\omega}+\Delta T_n^{2\omega}\\
T_f = T_f^{1\omega}+\Delta T_f^{2\omega} = T_f^{0}+\Delta T_f^{1\omega}+\Delta T_f^{2\omega}
\end{split}
\label{EquationS1}
\end{equation}

\noindent where $T_{n/f}$ is the actual temperature at the two sensors. $T_{n/f}^{1\omega}$ is the constant temperature composed of the initial temperature $T_{n/f}^{0}$ and the temperature rise $\Delta T_{n/f}^{1\omega}$, and $\Delta T_{n/f}^{2\omega}$ is the $2\omega$ temperature perturbation. Consequently, the corresponding resistances of the sensors are:

\begin{equation}
\begin{split}
R_n = R_n^{1\omega}+\Delta R_n^{2\omega} = R_n^{0}+\Delta R_n^{1\omega}+\Delta R_n^{2\omega}\\
R_f = R_f^{1\omega}+\Delta R_f^{2\omega} = R_f^{0}+\Delta R_f^{1\omega}+\Delta R_f^{2\omega}
\end{split}
\label{EquationS2}
\end{equation}

\noindent To be more detailed, the original $R-T$ curves of the two sensors are plotted by solid lines in Fig. \ref{Calibration_1}. By applying a 50 $\rm{\upmu A}$ d.c. current at the two sensors, we can measure the $2\omega$ resistance perturbation $\Delta R_{n/f}^{2\omega}$ and extract the corresponding $\Delta T_{n/f}^{2\omega}$ utilizing the original $R-T$ curves, as shown in Fig. \ref{Calibration_1} c-d. The raw data of the longitudinal and transverse $2\omega$ voltage are shown in Fig. \ref{Calibration_2}a-b. Therefore, we can calculate the thermoelectric coefficients \textit{via} the following equations:

\begin{equation}
\begin{split}
S_{xx} = -\frac{V_{xx}}{\Delta T_{n}^{2\omega}-\Delta T_{f}^{2\omega}}\cdot \frac{L_x}{L_0}\\
S_{xy} = -\frac{V_{xy}}{\Delta T_{n}^{2\omega}-\Delta T_{f}^{2\omega}}\cdot \frac{L_y}{L_0}
\end{split}
\label{EquationS3}
\end{equation}

\noindent The calculated thermoelectric coefficients are shown in Fig. \ref{Calibration_2}c-d. Nevertheless, it should be noted that the $T$-coordinate here is the readout temperature $T^0$ from our thermocouple rather than the actual temperature $T^{1\omega}$. To calibrate the $T^{1\omega}$, we measured the $R-T$ curves of the two sensors with the applied heating current, plotted as the dashed lines in Fig. \ref{Calibration_1}b. By comparing the solid curves with the dashed curves, we can extract the actual temperature $T_{n/f}^{1\omega}$ at the two sensors, respectively, as shown in Fig. \ref{Calibration_1}e-f. $T_{n}^{1\omega}$ and $T_{f}^{1\omega}$ exhibit slight differences, which means that the heating current also produces a constant temperature gradient across our sample. However, the constant signal wasn't taken into account in our $2\omega$ measurement. The actual temperature of our sample should be approximately derived as a linear combination of the temperature at the two sensors, $\alpha T_{n}^{1\omega}+(1-\alpha)T_{f}^{1\omega} (0<\alpha<1)$. Here we determined the value of $\alpha$ by comparing the inflection of the $S_{xx}$ and the $S_{xy}$ curves, which signifies the Néel temperature, with the $T_N$ determined by AHE. The ultimately calibrated $S_{xx}-T$ and $S_{xy}-T$ curves are presented in Fig. \ref{Calibration_2}e-f.

\clearpage
\newpage

\begin{figure*}[ht!]
	\centering
	\includegraphics[width=0.7\textwidth]{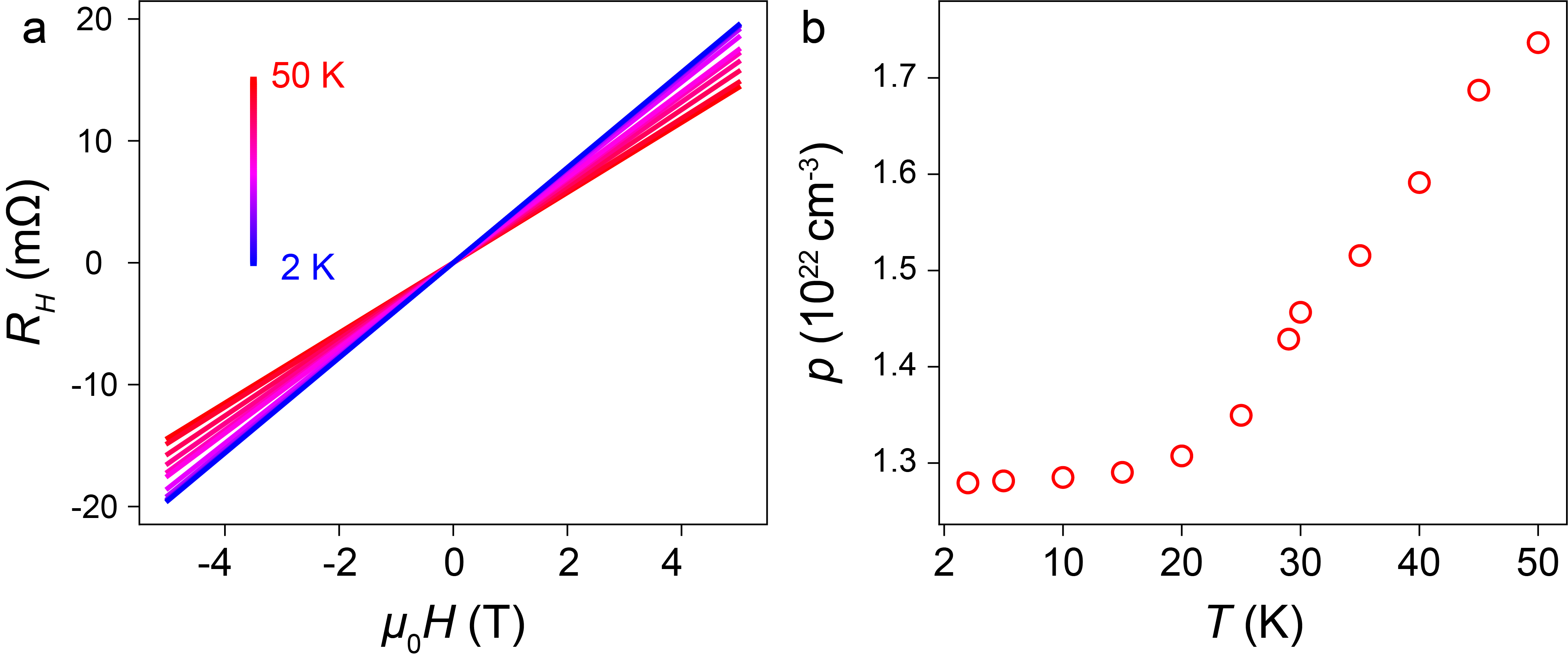}
	\caption{\textbf{Normal Hall effect and calculated carrier concentrations of device 1.} 
		\textbf{(a),} Normal Hall resistance extracted from the transverse voltage measurements at different temperatures. \textbf{(b),} Calculated hole concentration \textit{versus} temperature \textit{via} the equation $\frac{V_{H} t}{I B}=\frac{1}{p e}$, where $t$ is the thickness of our device (125 nm) and $p$ is the hole concentration.
	}
	
	\label{Normal_Hall}
\end{figure*}

\begin{figure*}[ht!]
	\centering
	\includegraphics[width=0.7\textwidth]{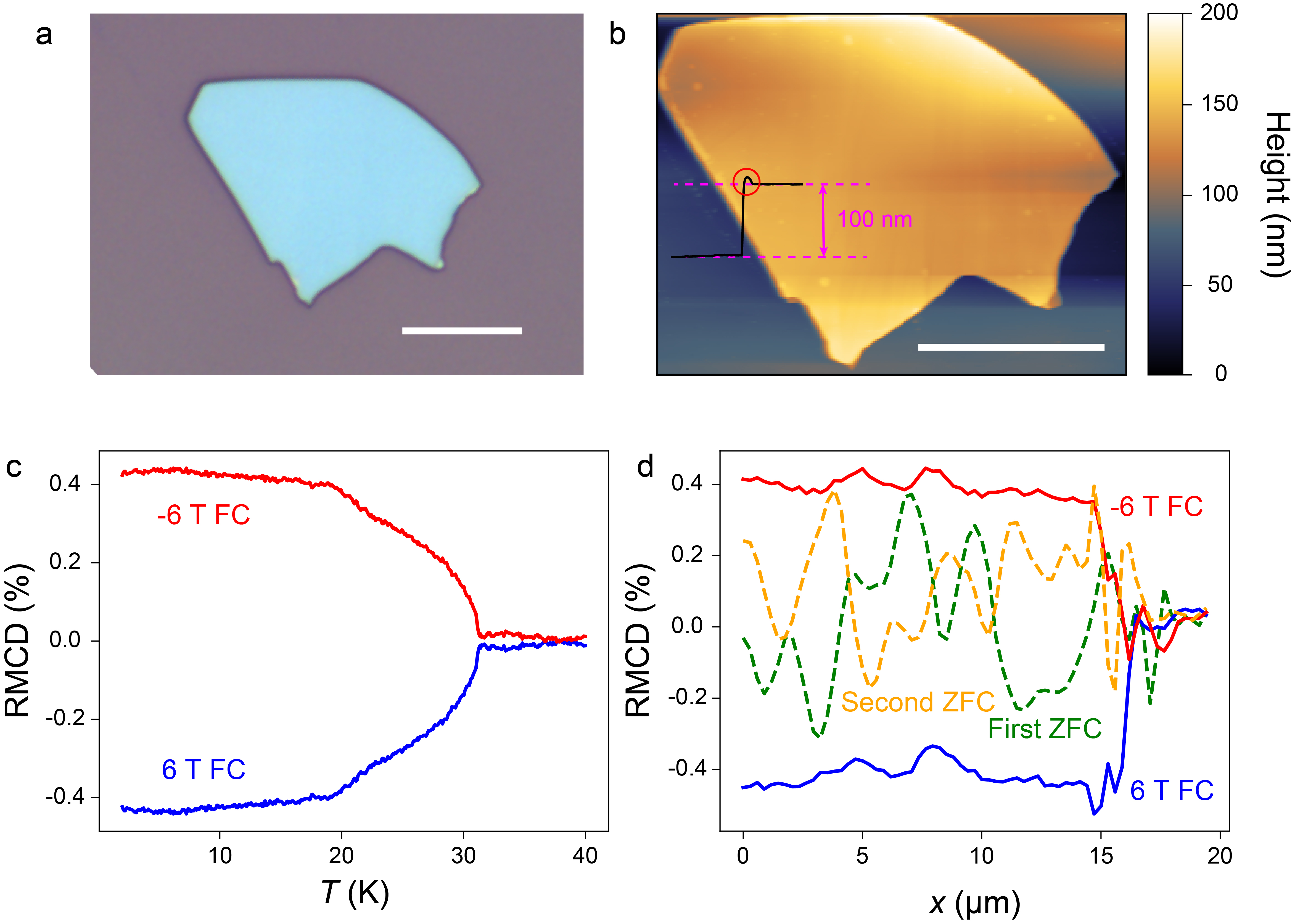}
	\caption{\textbf{Optical, atomic force microscopy and RMCD characterizations of the exfoliated $\rm{Co_{1/3}NbS_2}$ sample in Fig. 3 of the manuscript.} 
		\textbf{(a),} Optical image of the exfoliated sample. \textbf{(b),} Atomic force microscopy height image of the exfoliated sample which shows the thickness of $\sim$ 100 nm. The inset line profile shows a line cut from the atomic force microscopy  image, and the red circle labels the protrusion at the sample edge. Both of the white scale bars in (a) and (b) are 10 $\rm{\upmu m}$. \textbf{(c),} Single-point RMCD signal of the sample \textit{versus} temperature with 6 T and $-$6 T FC. The RMCD values are consistent with the mappings presented in Fig. 3 of the manuscript. \textbf{(d),} RMCD value \textit{versus} $x$-coordinate extracted from the line cut of Fig. 3. The ZFC domains exhibit similar RMCD values as the FC measurements, confirming the reliability of our experiments.
	}
	
	\label{RMCD_basic}
\end{figure*}

\clearpage
\newpage

\begin{figure*}[ht!]
	\centering
	\includegraphics[width=0.35\textwidth]{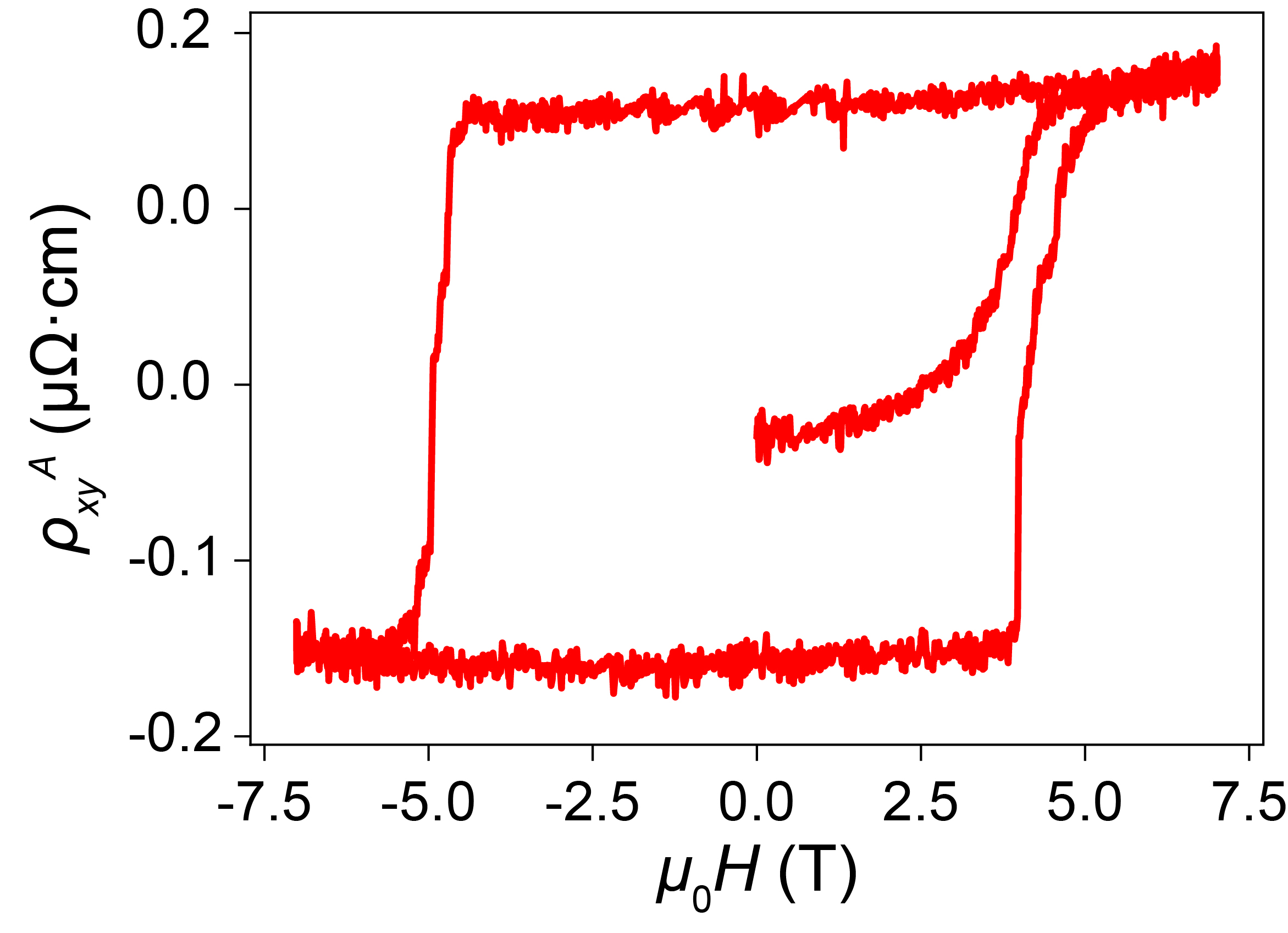}
	\caption{\textbf{AHE with the initial magnetization process of device 2 at 27.5 K.} 
	}
	
	\label{Initial_magnetization}
\end{figure*}

\begin{figure*}[ht!]
	\centering
	\includegraphics[width=1.0\textwidth]{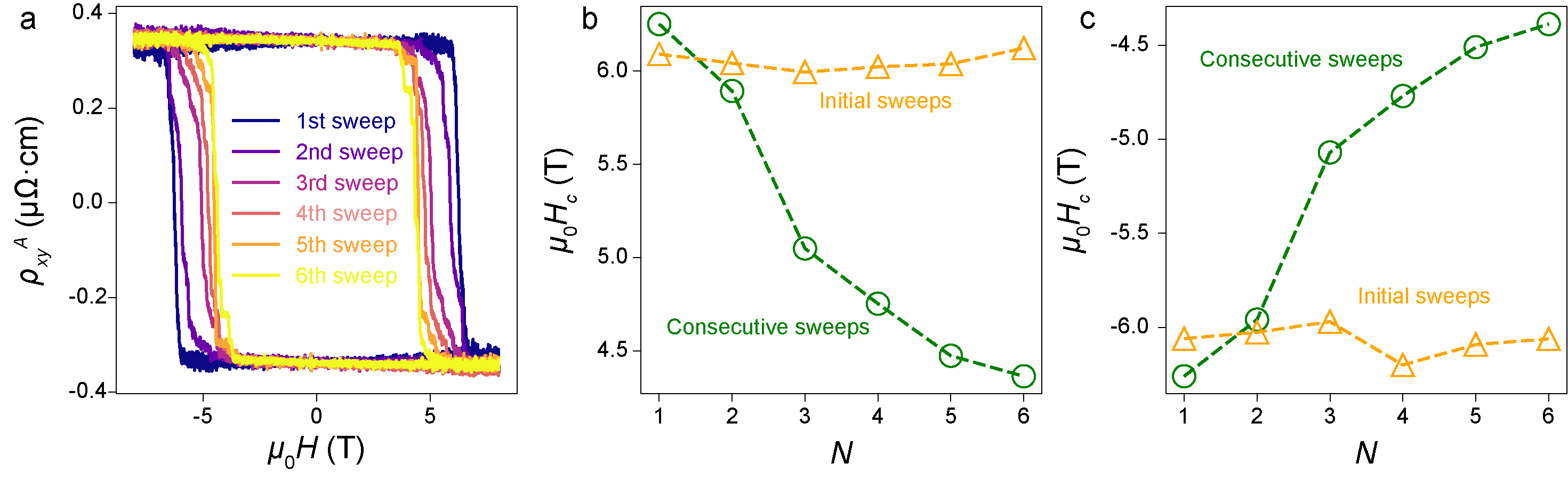}
	\caption{\textbf{The decrease of $H_c$ with consecutive magnetic field sweeps in device 1.} 
		\textbf{(a),} $\rho_{xy}^A$ \textit{versus} $\mu\rm{_0}$$H$ during several consecutive magnetic field sweeps at 27 K. \textbf{(b, c),} The green curves plot the extracted sweep-down (b) and sweep-up (a) $H_c$ \textit{versus} sweeping number $N$, while the orange curves plot the corresponding $H_c$ extracted from six initial sweeps after cooling down from 50 K. The $H_c$ of device 1 exhibits similar decreasing behavior upon consecutive sweeps as device 2, which is presented in Fig. 4c of the manuscript.
	}
	
	\label{Hc-N}
\end{figure*}

}
\bibliographystyle{naturemag}